\documentclass[twocolumn, superscriptaddress, prmaterials, amsmath,amssymb,noeprint]{revtex4-1}

\usepackage{hyperref}
\hypersetup{backref,colorlinks=true,allcolors=prussianblue}
\usepackage{xr}
\usepackage{graphicx}
\graphicspath{{./Figures-JPEG/}}
\usepackage[caption=false]{subfig}

\usepackage[table]{xcolor}
\definecolor{prussianblue}{rgb}{0.0, 0.19, 0.33}

\usepackage{setspace}
\usepackage{float}
\usepackage{indentfirst}

\usepackage{tcolorbox}
\usepackage{todonotes}

\usepackage{tikz}
\usetikzlibrary{calc,arrows,shapes,positioning}
\usetikzlibrary{patterns,snakes}
\usetikzlibrary{positioning,decorations.pathreplacing}

\definecolor{sangria}{rgb}{0.57, 0.0, 0.04}
\definecolor{arsenic}{rgb}{0.23, 0.27, 0.29}
\definecolor{prussianblue}{rgb}{0.0, 0.19, 0.33}
\definecolor{phthalogreen}{rgb}{0.07, 0.21, 0.14}
\definecolor{dgreen}{rgb}{0.0, 0.4, 0.2}

    \tikzstyle{blockg} = [rectangle, draw=arsenic!50, font=\fontsize{14pt}{14pt}\selectfont, minimum height=1cm, text centered,line width=1mm]    

    \tikzstyle{blockr} = [rectangle, draw=sangria!50, font=\fontsize{14pt}{14pt}\selectfont, minimum height=1cm, text centered,line width=1mm]    
    
    \tikzstyle{blockc} = [circle, draw=white, font=\fontsize{14pt}{14pt}\selectfont,
    text width=4em, text centered,line width=1mm]    

    \tikzstyle{line} = [draw=arsenic, -latex', line width=0.1cm]

\begin{document}

\author{Mohamed Hendy}
\affiliation{Department of Mechanical Engineering, University of British Columbia, 2054 - 6250 Applied Science Lane, Vancouver, BC, V6T 1Z4, Canada}
\author{Okan K. Orhan}
\affiliation{Department of Mechanical Engineering, University of British Columbia, 2054 - 6250 Applied Science Lane, Vancouver, BC, V6T 1Z4, Canada}
\author{Homin Shin}\affiliation{Quantum and Nanotechnologies Research Centre,
National Research Council Canada, Ottawa, ON K1N 5A2, Canada}
\author{Ali Malek}\affiliation{Clean Energy Innovation Research Centre, National Research Council Canada, Vancouver, BC V6T 1W5, Canada }
\author{Mauricio Ponga} \email[Corresponding author: ]{mponga@mech.ubc.ca}
\affiliation{Department of Mechanical Engineering, University of British Columbia, 2054 - 6250 Applied Science Lane, Vancouver, BC, V6T 1Z4, Canada}

\title{GAP-DFT: A graph-based alchemical perturbation density functional theory for catalytic high-entropy alloys}

\begin{abstract} 
High-entropy alloys (HEAs) exhibit exceptional catalytic performance due to their complex surface structures. However, the vast number of active binding sites in HEAs, as opposed to conventional alloys, presents a significant computational challenge in catalytic applications.  To tackle this challenge, robust methods must be developed to efficiently explore the configurational space of HEA catalysts. Here, we introduce a novel approach that combines alchemical perturbation density functional theory (APDFT) with a graph-based correction scheme to explore the binding energy landscape HEAs. Our results demonstrate that APDFT can accurately predict binding energies for isoelectronic permutations in HEAs at minimal computational cost, significantly accelerating configurational space sampling. However, APDFT errors increase substantially when permutations occur near binding sites. To address this issue, we developed a graph-based Gaussian process regression model to correct discrepancies between APDFT and conventional density functional theory values. Our approach enables the prediction of binding energies for hundreds of thousands of configurations with a mean average error of 30 meV, requiring a handful of ab initio simulations.
\end{abstract}

\maketitle

\section{Introduction}
Climate change is a pressing global issue, with the increasing carbon content in the atmosphere triggering a devastating global warming effect~\cite{adedeji2014global}. 
To mitigate this crisis and achieve net-zero emissions, an immediate transition to green energy sources is imperative. 
Catalytic materials are poised to play a critical role in this transition due to their potential to convert molecules into useful chemical compounds. %
However, the development of high-stable and active catalysts that can efficiently facilitate electrochemical reactions such as carbon dioxide reduction reaction (CO$_2$-RR)~\cite{yang2018design, johnson2021progress}, oxygen evolution reaction (OER)~\cite{suen2017electrocatalysis, mccrory2013benchmarking} and hydrogen evolution reaction (HER)~\cite{safizadeh2015electrocatalysis, dubouis2019hydrogen} remains challenging.
CO$_2$-RR has the potential to transform the production of carbon-based fuels to a closed carbon cycle with no net carbon emission through reducing CO$_2$ to fuels and chemicals.
Furthermore, OER and HER are the two main reactions for water splitting, which can be a crucial source of hydrogen production~\cite{you2018innovative,jamesh2016recent}. 
The widespread adoption of such green-energy approaches is highly dependent on developing suitable catalysts.

High-entropy alloys (HEAs) have shown tremendous potential for catalyzing various chemical reactions such as OER~\cite{hao2022unraveling, BATCHELOR2019834}, oxygen reduction reaction (ORR)~\cite{pedersen2020high, nellaiappan2020high} and CO$_2$-RR~\cite{nellaiappan2020high}, offering a promising solution to overcome the current limitations of traditional catalysts and accelerate the transition to a low-carbon future~\cite{li2021recent,katiyar2021perspective}. 
For instance, AuAgPtPdCu HEA was used to reduce carbon dioxide (CO$_2$) reaction to gaseous hydrocarbons \cite{pedersen2020high}. AuAgPtPdCu showed a higher selectivity towards complex hydrocarbon products at lower over-potential than pure Cu~\cite{nellaiappan2020high}.

This improved catalytic activity can be attributed to the vast number of active surface binding sites which arise due to all the possible combinations of the elements contained in the HEA as well as their tunable electronic structure~\cite{https://doi.org/10.1002/anie.202109212, PEDERSEN2021100651, https://doi.org/10.1002/adfm.202106715}.
Consequently, the HEA binding site activity is determined by atoms at the binding site and the atoms in the surrounding environment~\cite{saidi2022emergence}.
The different atomic combinations result in a continuous spectrum of binding energy due to the different perturbations around each binding site~\cite{BATCHELOR2019834, wang2021understanding}.
This unique and intrinsic feature of HEAs allows for tailoring the composition to optimize the catalytic reaction towards the reaction intermediates and break the scaling relationship~\cite{BATCHELOR2019834}.

From a computational viewpoint, the exceedingly large number of binding sites of HEAs poses a huge challenge for investigating HEAs for catalytic applications~\cite{BATCHELOR2019834, wang2021understanding}.
Hence, investigating just a single HEA composition is a nontrivial task that requires a large number of brute-force calculations and a high computational cost.
A common way to evaluate the catalytic activity is to use brute-force density functional theory (DFT) calculations to evaluate the binding energy (BE) as an activity descriptor~\cite{burke2012perspective,jones2015density}. 
Although brute-force DFT may be suitable for individual systems, it becomes prohibitively expensive when dealing with the large search space of HEAs, hence searching the entire materials space in this manner would be impractical and undesirable~\cite{li2021multi,yu2022high}.

To tackle this issue, robust approaches must be developed to efficiently screen the configurational space of catalytic HEA materials. 
One way to reduce the number of brute-force DFT calculations required to explore the compositional space is to use machine learning (ML) models. 
Even with ML models, thousands of computationally expensive DFT calculations are still required to achieve acceptable accuracy.
Previous work has used a feed-forward artificial neural network trained with $\sim1000$ DFT calculations of CO and OH adsorbates on bimetallic alloys, showcasing a root square mean error of $0.2$~eV~\cite{C7TA01812F}. {\color{black}Another work by Chen \emph{et al.} used more than $1000$ DFT calculations in a neural network model to study CO$_2$-RR on FeCoNiCuMo HEA surface with a mean average error of $0.095$ eV~\cite{chen2022machine}.}

However, the required training data increases with the complexity of the systems being studied, which is the case for HEAs.
For instance, Back \emph{et al.} used about $12,000$ DFT calculations to train a convolutional neural network (CNN) model to predict the BE of CO and H on intermetallic alloys with a mean average error (MAE) of $0.15$~eV~\cite{doi:10.1021/acs.jpclett.9b01428}. 
Furthermore, Pedersen \emph{et al.} employed an ML model for predicting the BE of CO on CuPdAgPtAu HEA using a smaller $2\times2$ supercell for training and validation of the ML model ~\cite{pedersen2020high}. This small cell can deem the DFT calculations inaccurate due to long-range electrostatic potential interactions of the adsorbate with its periodic image and its inability to represent the randomness of the HEA structures.  Indeed, when data obtained from a larger $3\times3$ cell was used, the ML model yielded an error twice as large when using a larger cell~\cite{pedersen2020high}. Hence, the large data sets needed to train ML pose a limitation on the size of the HEA system, which can reduce the prediction accuracy to larger cells. This issue shows the problems of building an ML model to explore HEAs due to their atomic complexity and extremely large compositional design space.

One efficient method to explore the compositional space is the alchemical perturbation density functional theory (APDFT), also known as computational alchemy~\cite{von2007alchemical, von2006molecular}.
APDFT relies on the concept of hypothetically varying the nuclear charge from a reference state to a target state.
The properties of the target system, such as energy, can be estimated relying on information from the reference system only, without the need to solve for the target system explicitly~\cite{von2007alchemical}.

APDFT has been mainly used to explore the compositional space of molecular systems.
The process of umbrella inversion of ammonia has been studied using APDFT to predict its energy barrier~\cite{sheppard2010alchemical}. 
Moreover, larger molecules such as fullerene structures have been explored using APDFT, with $C_{60}$ as a reference system.
Some carbon ($C$) atoms of $C_{60}$ has been transmuted to Boron ($B$) and Nitrogen ($N$) to explore the compositional space of $C_{x}(BN)_{y}$~\cite{balawender2018exploring}.

APDFT has recently been used to calculate some material properties for crystalline lattice systems.
It has been reported that APDFT qualitatively predicted the lattice stability, equilibrium volumes, and bulk moduli for many transition metals~\cite{to2016guiding}. 
The equilibrium volumes were predicted with less than $9\%$ error, while for the bulk moduli, the error was under $28\%$, indicating the need to assess the error for each material property.   
Some catalytic systems, such as carbides, nitrides, and oxides, have been studied using APDFT.
Moreover, APDFT yields reasonable accuracy for H, and OH adsorbate on TiC, TiO, and TiN \cite{griego2019benchmarking}. A limited number of studies explored the use of APDFT for catalysts pure metals doped with different elements~\cite{saravanan2017alchemical, https://doi.org/10.1002/qua.26380}. However, to the best of our knowledge, the application of APDFT to explore the catalytic properties of HEAs has not been explored. 

Here, we develop a graph-based alchemical perturbation density functional theory (GAP-DFT) to predict the BE of molecules in HEAs. First, we adapt and implement the APDFT approach to HEAs and asses its accuracy for permutations far and near the binding site. We show that for isoelectronic (single and double) swaps, the errors are deemed acceptable for three molecules, e.g., CO, O and H. However, we found that the errors are exceedingly large for binding site swaps, compromising the predictions of APDFT. Hence, we develop a graph-based Gaussian process regression (GPR) model to correct for BE errors near the binding sites. We show that the combination of APDFT and the graph-based approach can efficiently predict the BE energy of HEAs at an insignificant computational cost using a few \emph{ab initio} simulations. This approach can be used to explore the BE distributions while retaining sufficient accuracy for swaps near the binding site, opening up the possibility of efficiently exploring the design of HEAs for optimal catalytical performance. 

\section{Theoretical and computational methodology}\label{sec2}

In this section, we present the theoretical foundation of APDFT and provide a detailed overview of the computational methods employed in this study.

\subsection{APDFT theoretical background}\label{subsec1}

The general idea behind APDFT for catalysis problems relies on introducing small perturbations to the atomic charges from a reference system, with the constraint of keeping the transformation isoelectronic (i.e., the number of electrons are identical for the reference and target states)~\cite{https://doi.org/10.1002/qua.26380}.
When dealing with an HEA system with multiple elements, it is crucial to preserve the alloy's composition throughout the system.
We achieved this by permuting a pair of atoms with $\Delta$Z$=\pm1$ to get different perturbed structures, with Z the atomic number of the element.
This permutation strategy aims to preserve the number of electrons upon transmutation and the composition of the alloy under investigation, and it is well-suited to the HEA system.

We conducted three DFT calculations for the reference system on the HEA surface, an adsorbate molecule, and a surface with the adsorbate. The BE was then calculated according to the following equation~\cite{chen2018atomic}

\begin{align}\label{eq:eq1}
\mathrm{BE} = \mathrm{E}_\mathrm{surf} + \mathrm{E}_\mathrm{ads} - \mathrm{E}_\mathrm{ads-surf}
\end{align}

where $\mathrm{E}_\mathrm{surf}$ is the energy of a system with a pristine surface, $\mathrm{E}_\mathrm{ads}$ is the energy of the adsorbate and $\mathrm{E}_\mathrm{ads-surf}$ is the energy of a system consisting a surface with an adsorbate. With the definition in Eq.~\ref{eq:eq1}, positive BE values indicate {\color{black}stables intermediates, while negative unstable ones. }

\begin{figure}[t!]
\centering
\includegraphics[trim={0cm  0.2cm 0cm 0cm },clip,width=1\columnwidth]{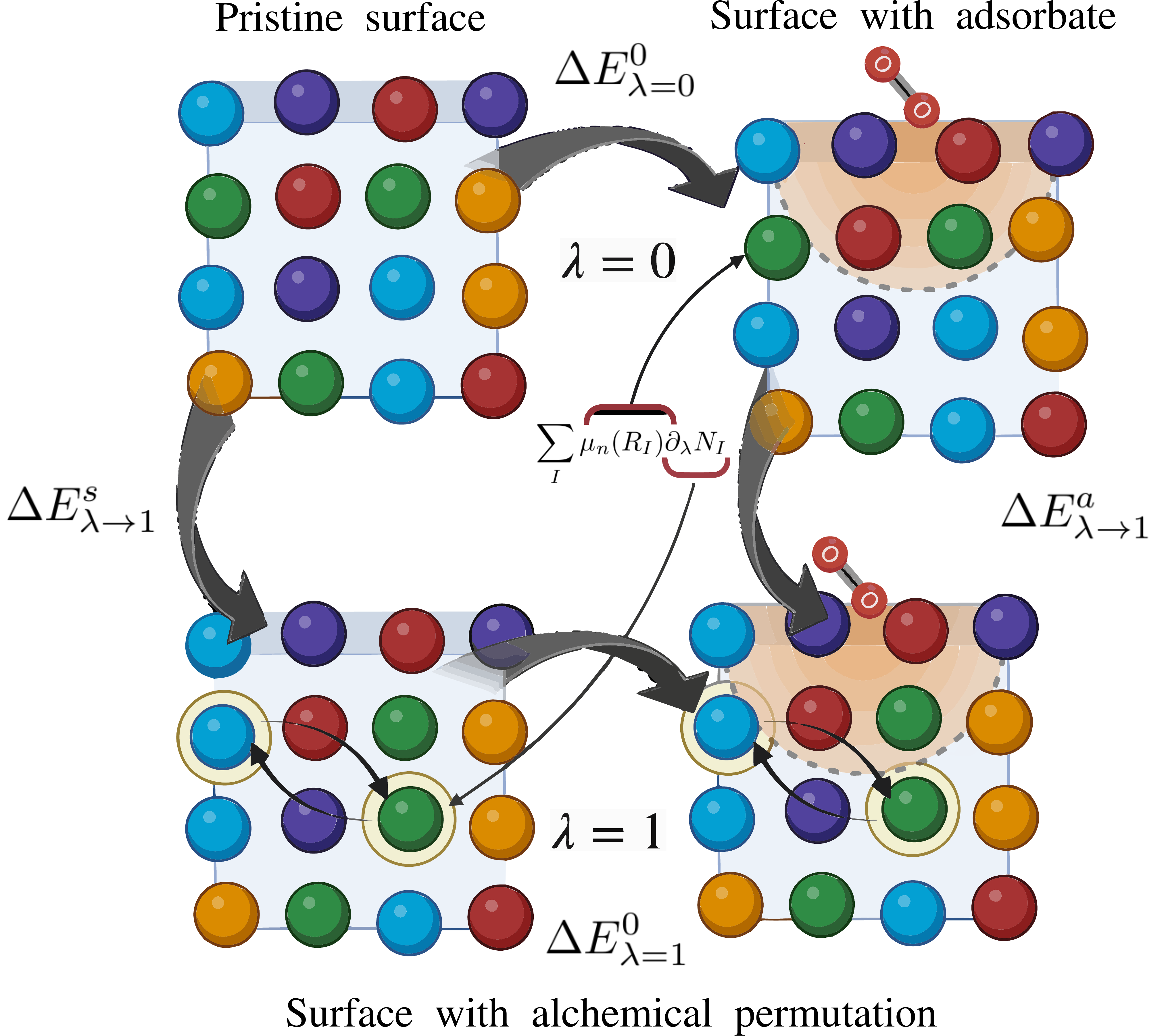}
\caption{Thermodynamic cycle: Reference state ($\lambda=0$) to permuted (swapped) state ($\lambda=1$) for a HEA, where $\Delta E_{\lambda=0}^0$  is the energy difference between the reference state with and without adsorbate and it is related to the BE calculated using DFT according to Eq.~\ref{eq:eq1}, $\Delta E_{\lambda\rightarrow1}^a$ is energy change due to swapping two atoms in the surface with adsorbate, $\Delta E_{\lambda\rightarrow1}^s$ is similar to $\Delta E_{\lambda\rightarrow1}^a$ but for the surface slab without adsorbate and $\Delta E_{\lambda=1}^0$ is energy difference between the permuted system with and without adsorbate.
}
\label{Fig:alchemy_schematic}
\end{figure}

Here is how APDFT can be used to estimate the BE of multiple target systems.
Let $\lambda$ denote the reaction path, and the reference system is labelled with $\lambda=0$, while the target {\color{black}perturbed} system is $\lambda=1$, as shown in Fig.~\ref{Fig:alchemy_schematic}. Then, using the thermodynamic cycle between the reference and target system in  Fig.~\ref{Fig:alchemy_schematic}, we can get~\cite{https://doi.org/10.1002/qua.26380}
\begin{align}\label{eq:eq2}
\Delta E_{\lambda=0}^0 + \Delta E_{\lambda\rightarrow1}^a = \Delta E_{\lambda=1}^0 + \Delta E_{\lambda\rightarrow1}^s
\end{align}
where $\Delta E_{\lambda=0}^0$ is the difference in energy between the reference state with and without adsorbate and it is related to the BE calculated through DFT according to Eq.~\ref{eq:eq1}, $\Delta E_{\lambda\rightarrow1}^a$ is the change in energy due to swapping atoms in the surface plus adsorbate, $\Delta E_{\lambda\rightarrow1}^s$ is similar to $\Delta E_{\lambda\rightarrow1}^a$ but for the pristine surface slab and $\Delta E_{\lambda=1}^0$ is the difference in energy between the target system with and without adsorbate. By rearranging  Eq.~\ref{eq:eq2}, we get
\begin{align}\label{eq:eq3}
\Delta BE = \Delta E_{\lambda=1}^0 - \Delta E_{\lambda=0}^0 = \Delta E_{\lambda\rightarrow1}^a - \Delta E_{\lambda\rightarrow1}^s
\end{align}
note that the difference between $\Delta E_{\lambda=1}^0$ and $\Delta E_{\lambda=0}^0$ is the difference in the BE between the swapped state $\lambda=1$ and the reference state $\lambda=0$.

Then, the energy of the target state can be expanded using Taylor's expansion as a function of the energy of the reference state with some additional terms known as alchemical derivatives as~\cite{sheppard2010alchemical}
\begin{align}\label{eq:taylor}
\Delta E_{\lambda=1}^0 = \Delta E_{\lambda=0}^0 + \partial_{\lambda} E^0 \Delta\lambda + \frac{1}{2} \partial_{\lambda}^2 E^0 \Delta\lambda^2 +\text{H.O.T.},
\end{align}
where $\Delta$ $E_{\lambda=1}^0$ and $\Delta$ $E_{\lambda=0}^0$ are the energies of the target and reference states, respectively. $\partial_{\lambda} E^0$ denotes the derivative of $E_{\lambda=0}^0$ with respect to $\lambda$, and \text{H.O.T.} denotes the higher-order terms. Assuming that the changes between the reference and target states are isoelectronic with the number of electrons of both systems being the same, then~\cite{balawender2018exploring}
\begin{align}\label{eq:alchderv}
\partial_{\lambda} E^0 = \sum_I \mu_n(R_I) \partial_{\lambda} N_I.
\end{align}
In Eq.~\ref{eq:alchderv}, $\mu_n(R_I)$ denotes the electrostatic potential of the reference-system's atoms located at positions $R_I$, while $\partial_{\lambda} N_I$ denotes the difference in nuclear charges between the reference and target systems at the nuclear positions. This term accounts for the nuclear chemical potential gradient $\mu_n(R_I)$ due to variation in nuclear charges $N_I$ at the positions $R_I$ from the reference to the target states. The electrostatic potential ($\mu_n(R_I)$) has a nuclear and electronic components and can be calculated as~\cite{balawender2018exploring}
\begin{equation}\label{Eq:firstAlchDrvnn}
 \mu_n(R_I) =\sum_{B\neq A} \frac{Z_B}{|R_A-R_B|}-\int \frac{\rho}{|r-R_A|}dr,
\end{equation}
where the first term represents the electrostatic potential of all the nuclear charges felt by each atom, while the second term describes the electrostatic potential of the electrons felt by each atom. In summary, the APDFT predictions are grounded in the electrostatic potential information from the reference system's DFT calculations and the atomic number difference between swapped atoms, which depends on each specific perturbed target system.

\subsection{Computational details of DFT calculations}\label{subsec2}

The Vienna Ab initio Simulation Package (VASP) ~\cite{PhysRevB.49.14251,PhysRevB.54.11169} was used to conduct the Kohn-Sham DFT calculations. The projector augmented wave method (PAW) was employed with ultra-soft pseudo-potentials. The generalized gradient approximation (GGA) with the revised Perdew-Burke-Ernzerh (RPBE)~\cite{PhysRevB.59.7413} exchange-correlation functional was chosen. A plane-wave basis set with an energy cut-off of $700$~eV was used. The force on each ion for the ionic relaxation calculations was less than $0.05$~eV$\cdot$\r{A}$^{-1}$. 

The initial lattice parameter of the CuPdAgPtAu HEA was taken as the average of the lattice parameter of each element in bulk. After relaxation, the lattice parameter was $4.012$~\r{A}. The initial supercell was created to represent a [$111$]-face-centred cubic (FCC) surface with the stacking of ABC. The supercell was $3\times3$ of $5$ layers consisting of $45$ atoms of equimolar fraction for each element. A vacuum of $18$~\r{A} was inserted in the direction perpendicular to the surface to avoid surface-surface interaction. A Monkhorst-Pack $k-$points of $6\times6\times1$ were used, where the gamma point was used along the [$111$] direction.

We implemented the APDFT method using a set of Python and bash scripts on HEA, specifically a 45-atom equiatomic CuPdAgPtAu HEA supercell. The main input was the atomic-centred electrostatic potential from the VASP output file, which is used to calculate the alchemical derivative from Eq.~\ref{eq:alchderv}, with $\partial_{\lambda} N_I$ as the other input, which depends on the difference in the nuclear charge of the reference system and the new perturbed system. The BE of the perturbed systems were calculated as the summation of the BE of the reference system from Eq.~\ref{eq:eq1} and the alchemical derivative from Eq.~\ref{eq:alchderv}. The implementation was tested, and the capabilities of APDFT were assessed using three different adsorbates: carbon monoxide (CO), oxygen (O) atom, and hydrogen (H) atom. The CO molecule was placed in the on-top binding site, while the O and H atoms were placed in the hexagonal close-packed (HCP) hollow binding site. Full relaxation of the molecule/atoms was performed before doing the APDFT scheme. The codes used for the APDFT calculations are available at \cite{hea-apdft-website}.

\section{Results}\label{sec3}

\subsection{APDFT errors assessment for HEA}

We first performed three calculations of the HEA surface slab, the adsorbate molecule/atom, and a slab with the adsorbate molecule/atom. These calculations represent the reference system calculations from which the APDFT estimations were calculated at negligible additional computational cost.
Using this single DFT reference system, we were able to predict the BE of thousands of perturbed systems with negligible additional computational cost.
We limited the swaps to three different types with $\Delta$Z$=\pm1$: Pt with Au (Pt~$\leftrightarrow$~Au), Pd with Ag (Pd~$\leftrightarrow$~Ag) and the simultaneous swaps of Pt with Au and Pd with Ag (Pt~$\leftrightarrow$~Au~$+$~Pd~$\leftrightarrow$~Ag).
The total combinations number of perturbed systems resulting from these swaps is 81 (9 $\times$ 9) for Pt~$\leftrightarrow$~Au or Pd~$\leftrightarrow$~Ag, and 6561 (9 $\times$ 9 $\times$ 9 $\times$ 9) for Pt~$\leftrightarrow$~Au~$+$~Pd~$\leftrightarrow$~Ag.

To evaluate the accuracy of APDFT predictions, we performed 100 DFT calculations, including 90 for single-atom swaps and 10 for double-atom swaps for each molecule/atom.
The swaps were categorized into two groups. Those involving atoms on the direct binding site are referred to as \emph{binding site atom(s)}, whereas the remaining cases are labelled as \emph{non-binding site atoms}. The non-binding site atom cases include surface atoms, atoms in the bulk, and atoms surrounding the binding sites. 

\begin{figure}[H]
\centering
\subfloat[\centering CO on-top]{\includegraphics[trim={0cm  4cm 0cm 0cm },clip,width=0.9\columnwidth]{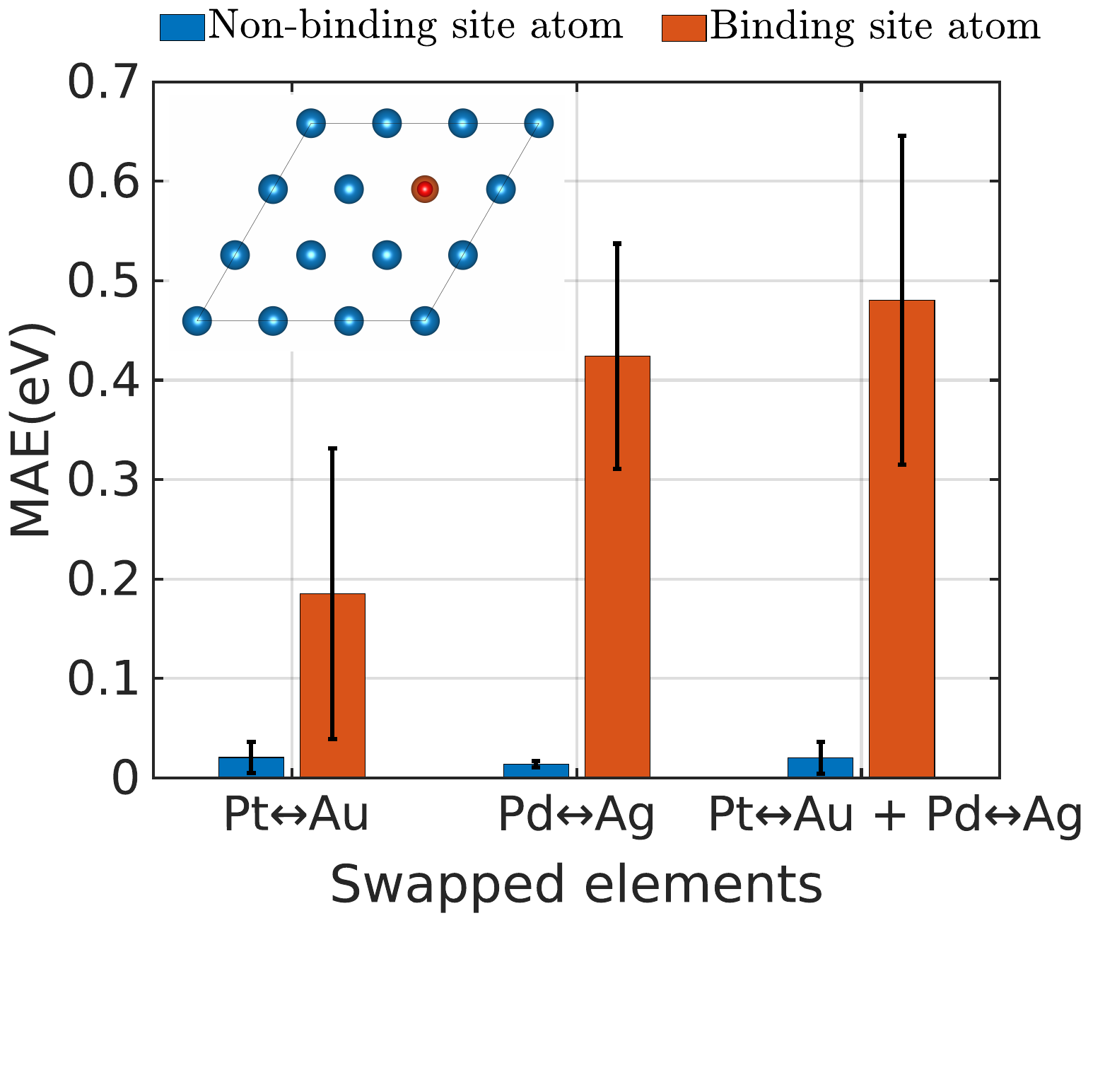}
}\hfill
\subfloat[\centering O hollow]{\includegraphics[trim={0cm  4cm 0cm 0cm }, width=0.9\columnwidth]{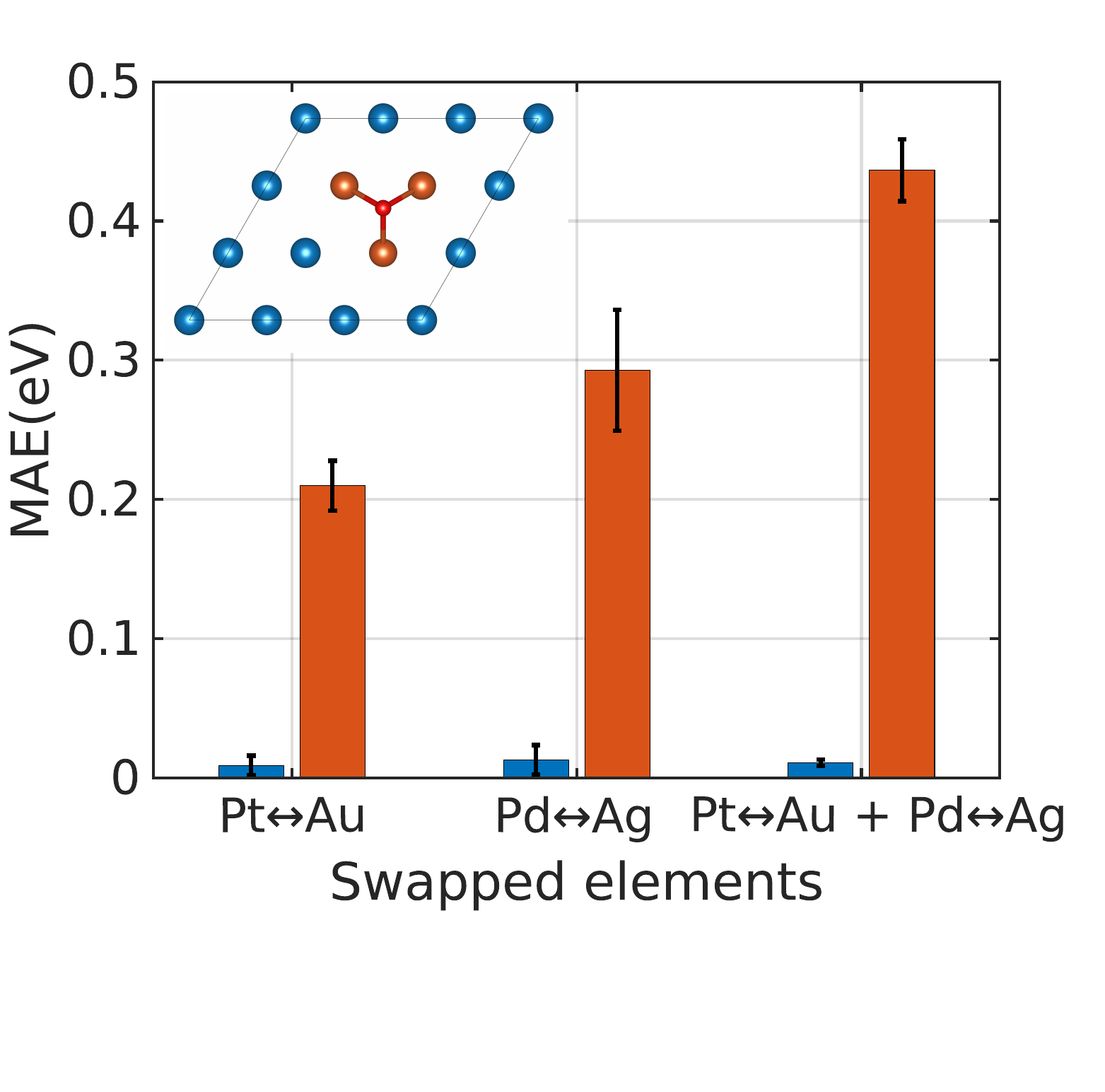}
}\hfill
\subfloat[\centering H hollow]{\includegraphics[trim={0cm  4cm 0cm 0cm },clip,width=0.9\columnwidth]{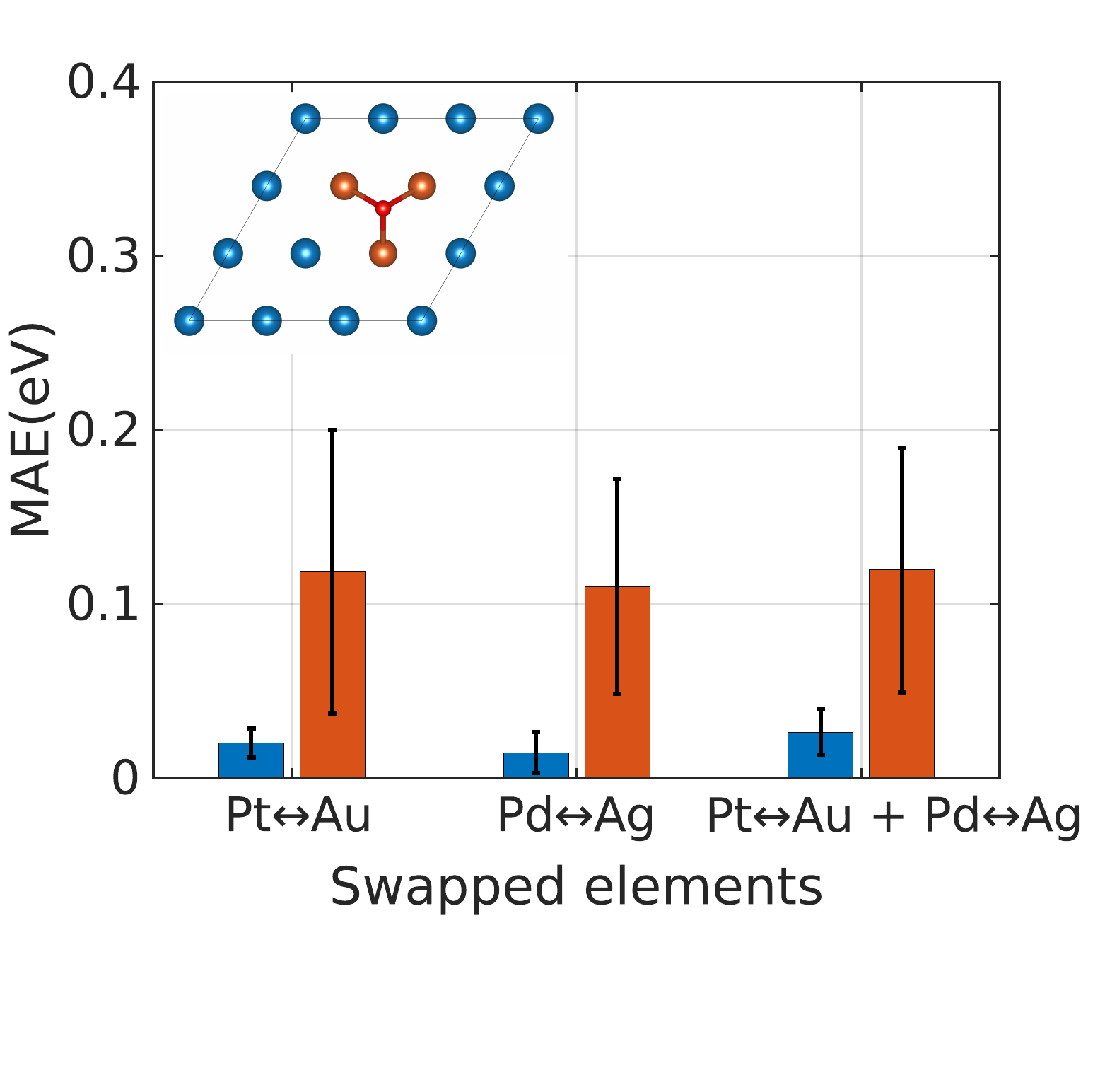}}
\caption{The mean average error (MAE) for each swapping case.
}
\label{Fig:hcp_swap}
\end{figure}

Fig.~\ref{Fig:hcp_swap} shows the mean average error (MAE) for three types of atom swapping: Pt~$\leftrightarrow$~Au, Pd~$\leftrightarrow$~Ag, and simultaneous Pt~$\leftrightarrow$~Au~$+$~Pd~$\leftrightarrow$~Ag swaps for the three different adsorbates, e.g., CO, O, and H. The inset of Fig.~\ref{Fig:hcp_swap} visually differentiates the binding site atoms, shown in orange, from other atoms, depicted in cyan, highlighting CO in the on-top site, while O and H in the HCP hollow site. {\color{black}We focused on these adsorption sites since they are the most favourable site for these adsorbates on transition metal surfaces~\cite{zhang2001density, chen2018atomic, ferrin2012hydrogen, xing2021adsorption}.} This comparison allows us to assess the impact of these specific atomic swaps on the BE.  

The results showcased in Fig.~\ref{Fig:hcp_swap} illustrate the higher accuracy of APDFT when used to permute non-binding site atoms as opposed to binding site atoms across all absorbates. More speciﬁcally, the mean average error (MAE) is less than 0.03 eV for the non-binding site atoms swapping for all adsorbates, as shown in Fig.~\ref{Fig:hcp_swap}. This precision in handling two simultaneous permuted atoms greatly enhances the predictive capability of APDFT, allowing for accurate simulations of a larger number of target systems. 

Another important observation from Fig.~\ref{Fig:hcp_swap} is that the error when swapping the binding site atoms is high for all permutations and adsorbates.
For CO, an MAE ~$\approx0.2$~eV was observed for Pt~$\leftrightarrow$~Au swapping, which is about ten times the error for swapping the non-binding site atoms.

A similar MAE was observed for O.  
This error reaches around $0.4$~eV for Pd~$\leftrightarrow$~Ag swapping with CO and $0.3$~eV with O, indicating that the error is dependent on the elements permuted. A higher error was recorded with Pd~$\leftrightarrow$~Ag permutation. 
Furthermore, the error is the highest for the double swap Pt~$\leftrightarrow$~Au~$+$~Pd~$\leftrightarrow$~Ag of around $0.5$~eV with CO and $0.45$~eV with O due to swapping two atoms at the same time in direct proximity to the adsorbate.
For H, the binding site atoms permutation for all three cases is slightly higher than $0.1$~eV. 
This error can be related to the large change in electron density in the proximity of the adsorbate, which APDFT failed to predict. {\color{black}Furthermore, the change in bond distance between the adsorbate and the binding site atom(s) is not considered in APDFT.}

\begin{figure*}
\centering
{\includegraphics[width=0.8\textwidth]{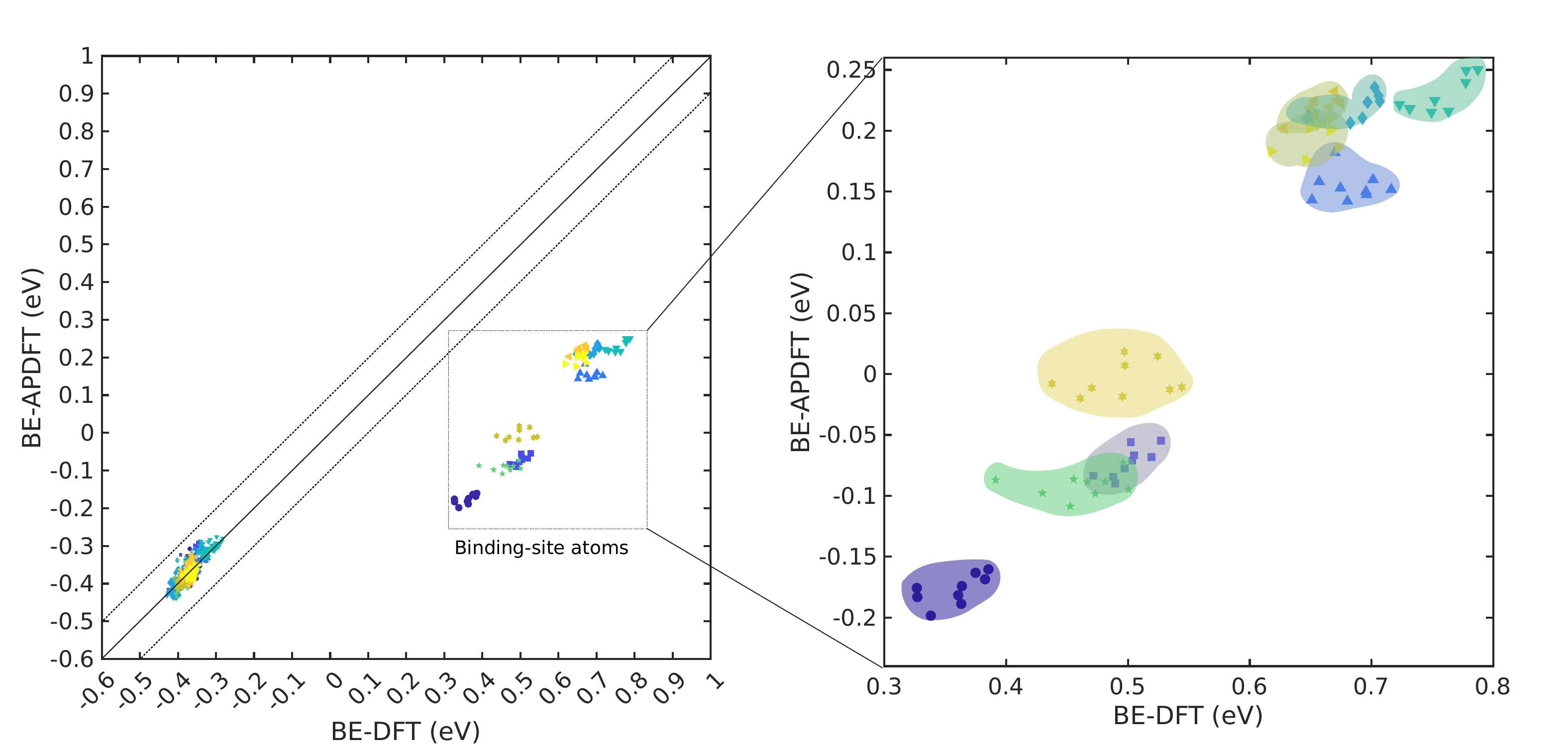}}
\caption{On the left, the binding energy (BE) in eV for swapping Ag with Pd on the binding site and non-binding atoms for nine different HEA configurations. The dotted lines represent a $\pm 0.1$~eV deviation from the DFT values. The binding site atoms swapping for each configuration are grouped on the right.
}
\label{Fig:HEA_conf_err}
\end{figure*}

To explore how different atomic environments influence APDFT errors, nine different HEA configurations were generated with CO on-top of an Ag atom. The BE of APDFT compared to DFT is shown in Fig.~\ref{Fig:HEA_conf_err}(a). The {\color{black}non-binding site atoms} swapped showed an error of less than $0.1$~eV and clustered together near the left bottom of the plot, indicating a small range of BE difference when swapping atoms away from the binding site atoms. On the other hand, the swapped binding site atoms have shown a wider range of BE with higher APDFT errors. Moreover, for each configuration, the swapping of the binding site Ag atom was done with all of the Pd atoms in the simulation cell, which tend to cluster together as shown in the right part of Fig.~\ref{Fig:HEA_conf_err}(a) in different colors. 

\begin{figure*}
\centering
\subfloat[\centering Pd to Ag]{\includegraphics[width=0.45\textwidth]{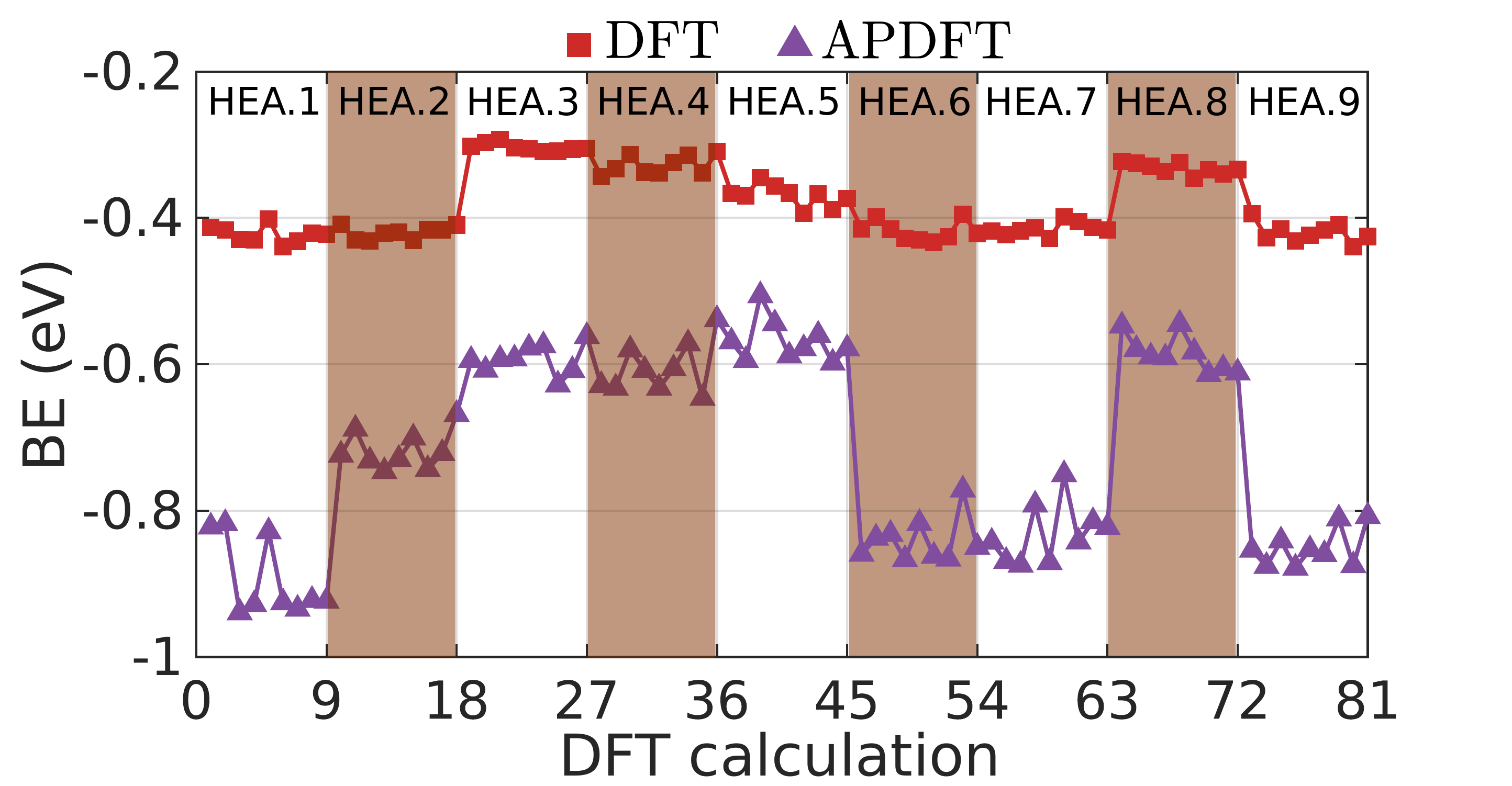}}
\subfloat[\centering Ag to Pd]{\includegraphics[width=0.45\textwidth]{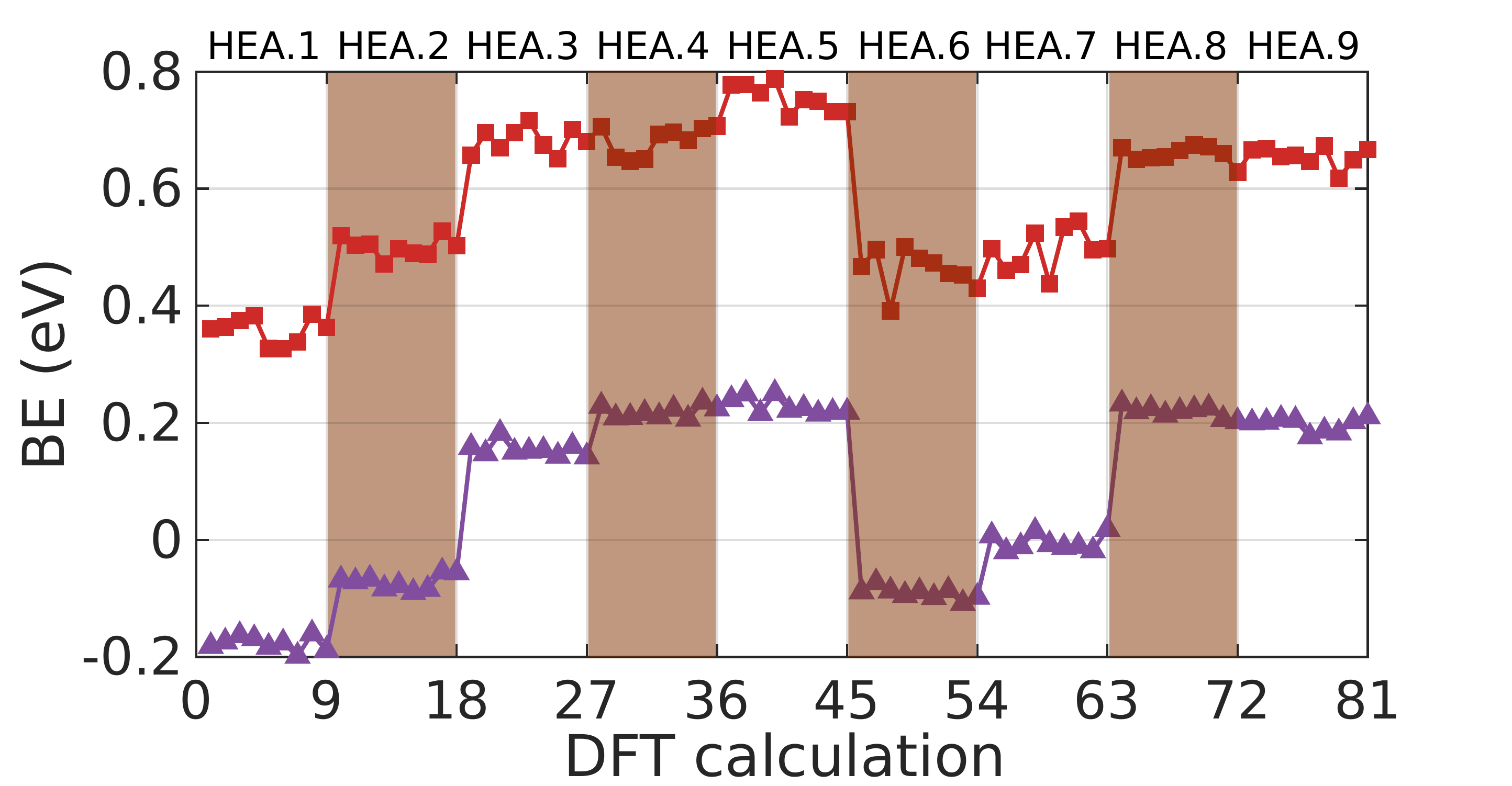}}
\bigbreak
\subfloat[\centering Pt to Au]{\includegraphics[width=0.45\textwidth]{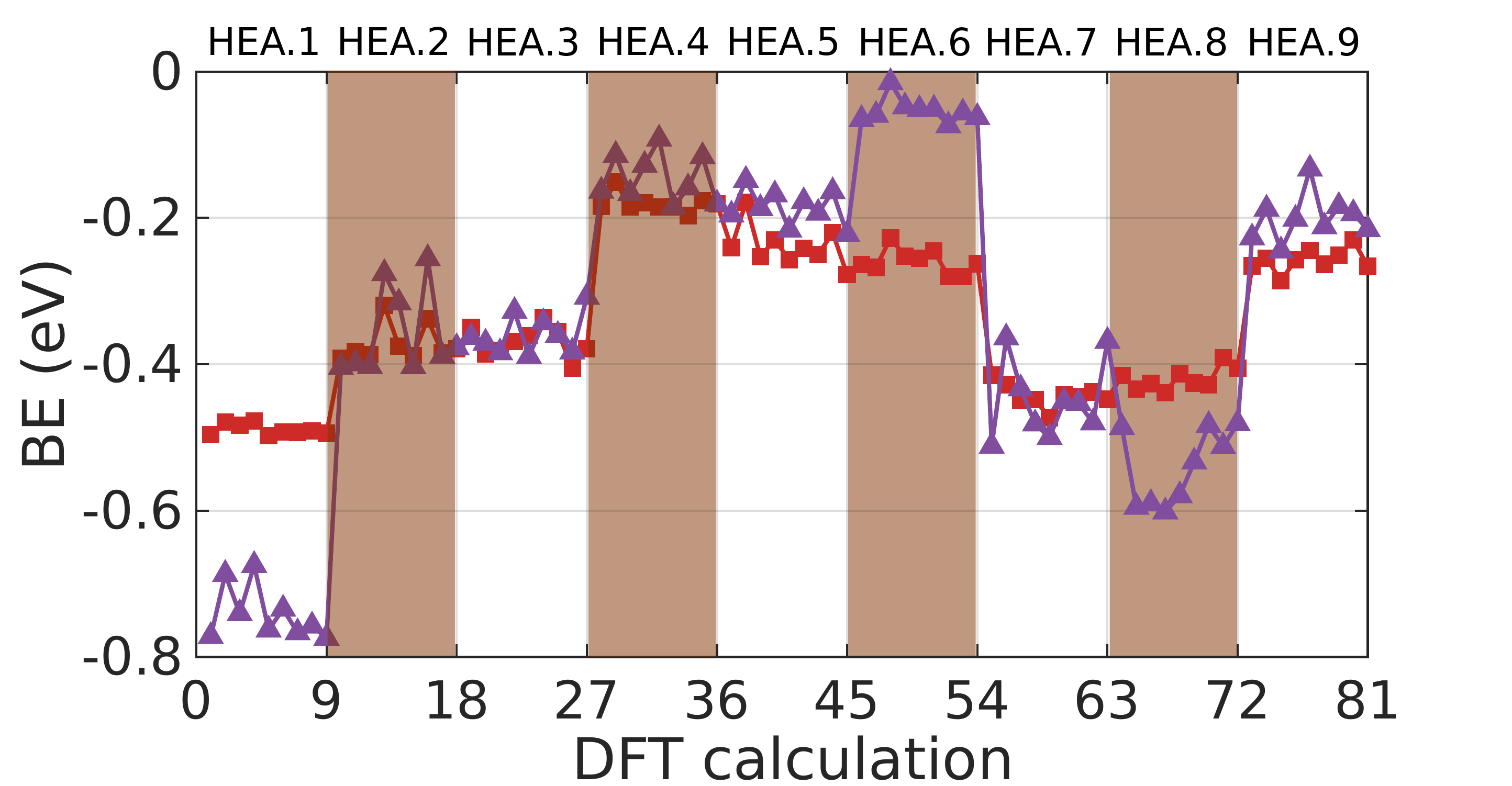}}
\subfloat[\centering Au to Pt]{\includegraphics[width=0.45\textwidth]{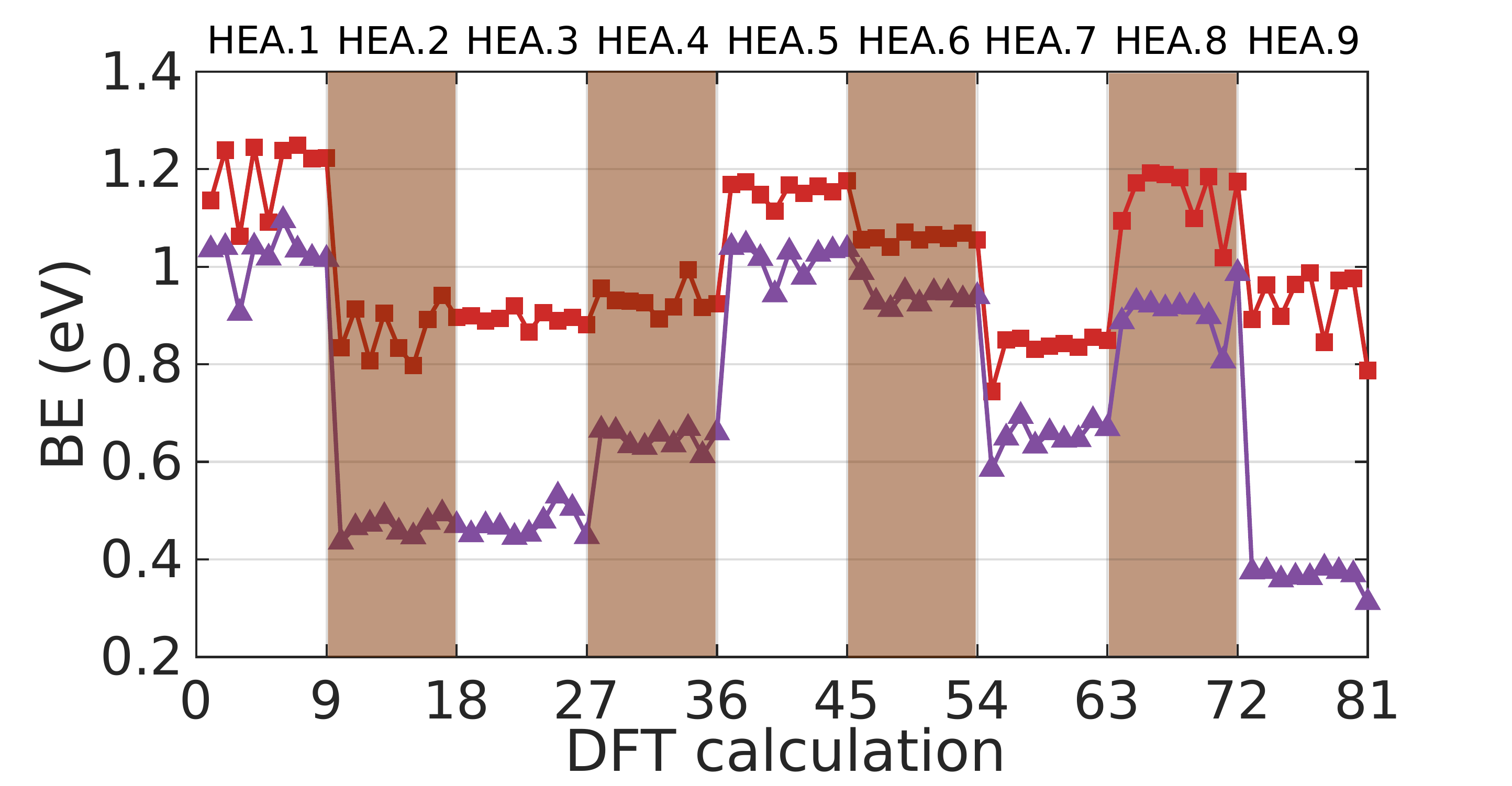}}
\caption{ The DFT and APDFT binding energy (BE) of CO on-top of an atom. A permutation of the binding site atom from (a) Pd to Ag, (b) Ag to Pd, (c) Pt to Au and (d) Au to Pt. 
}
\label{Fig:HEA_conf_DFT_alch}
\end{figure*}

To further quantify the errors, all the permutations involving the binding site atom were considered: Pd~$\leftrightarrow$~Ag, Ag~$\leftrightarrow$~Pd, Pt~$\leftrightarrow$~Au and Au~$\leftrightarrow$~Pt. Nine different HEA configurations were created for each of the four mentioned cases. For each of the nine configurations, nine swaps between the binding site atom and all the other atoms in the cell with $\Delta$Z$=\pm1$ were considered, and DFT calculations were performed and compared with the APDFT predictions as shown in Fig~\ref{Fig:HEA_conf_DFT_alch}. It can be observed that even though the errors are large, APDFT can {\color{black}\emph{qualitatively}} predict the BE changes with different HEA configurations {\color{black}for swaps involving Ag~$\leftrightarrow$~Pd and Pd~$\leftrightarrow$~Ag. When swaps involved Pt~$\leftrightarrow$~Au and Au~$\leftrightarrow$~Pt, the APDFT results showcase large errors and no qualitative behavior compared to DFT.} 

The large errors are attributed to quantum mechanical effects, which are ignored in the APDFT, and the inherent surface complexity of HEAs. In some cases, this error can affect the conclusion about whether a binding site is strong or weak binding to CO as in Fig~\ref{Fig:HEA_conf_DFT_alch}(b), which has positive BE values predicted by DFT; however, negative BE values predicted by APDFT. The error ranges between $0$~eV to $0.6$~eV, indicating that the error depends on the different atomic environments and the swapped atoms, which is difficult to model using analytical models. 
 
\subsection*{Graph-based correction for binding site energies}
After determining the potential and limitations of using APDFT to predict the BE, we identified a significant challenge when permuting atoms at the binding site. 
These errors are primarily attributed to non-classical quantum effects, which are not captured by APDFT's electrostatic considerations {\color{black} as well as the higher order terms of APDFT}. To address this issue without incurring additional computational costs, we developed a data-driven correction model that accounts for the complex interplay of spatial and chemical dependencies in the atomic environment.

The correction model, named GAP-DFT, is built on graph theory, ensures translation, rotation and permutation invariance relying on a modest dataset of DFT calculations. Next, we will describe the developed approach. Let $r_{ij}$ denote the Euclidean distance between atoms $i$ and $j$. $Z_i$ denotes the atomic number atom $i$. With these parameters, we can define arrays that describe the chemo-spatial environment of a binding site. For instance, the Coulomb matrix~\cite{rupp2012fast}, {\bf C} can be described by

\begin{align} 
C_{ij}= 
\begin{cases}
0.5Z_i^{2.4} & \text{ if } i=j \\
\frac{Z_i Z_j}{r_{ij}} \displaystyle & \text{ if } i \neq j.
\end{cases}
\end{align}

The diagonal elements of the Coulomb matrix represent the potential of the free atom approximated by a polynomial fit of the atomic energies to the nuclear charge $Z_i$. The off-diagonal terms capture the Coulomb repulsion between nuclei $i$ and $j$. It is worth mentioning that other similar arrays can be defined using a combination of electronegativity and electron affinity instead of $Z_i$. In analogy to the graph theory, {\bf C} here represents the adjacency matrix between the atoms in the system, and it incorporates the chemical and spatial information of the atoms. 

While {\bf C} ensures translation and rotation invariance, it lacks permutation invariance, meaning that different atom labelling for the same system can yield distinct Coulomb matrices. This limitation renders the Coulomb matrix, in its current form, a non-unique descriptor of the chemo-spatial environment in HEAs. Hence, the normalized Laplacian matrix can be defined as $\bf L = I - D^{1/2} \cdot C \cdot D^{1/2}$. ${\bf D}$ represents the degree matrix of C, as,

\begin{equation}
 {\displaystyle D_{ij}:=\left\{{\begin{matrix} \deg(C_{i})&{\mbox{if}}\ i=j\\0&{\mbox{otherwise}}\end{matrix}}\right.}
\end{equation}
where $ \deg(C_{i})$ denotes the sum of all components in the $i-$th column of {\bf C}. It is well known that the normalized Laplacian matrix is definitive positive and has real and bounded eigenvalues $\lambda_1 \leq \lambda_2 \ldots \leq \lambda_N$, where $N$ is {\color{black}the number of atoms in the super-cell}. As such, ${\bf L}$ can be decomposed using the spectral theorem as~\cite{das2004laplacian}

\begin{equation} \label{eq:spectral_decomp}
{\bf L} = \sum_{i=1}^{N} \mathbf U \Lambda U^T,
\end{equation}
where $\bf U$ is a maxtix and ${\boldsymbol \Lambda}$ a diagonal matrix containing the right eigenvectors and eigenvalues of ${\bf L}$, respectively.  Assume a chemical environment around an adsorbate. This configuration will naturally lead to a normalized Laplacian matrix ${\bf L}_a$. If one atom is permuted, this will lead to another normalized Laplacian matrix ${\bf L}_b$, which in principle will have different components than ${\bf L}_a$. Furthermore, the spectral decomposition in Eq.~\ref{eq:spectral_decomp} allows us to determine whether the configurations are identical by examining the eigenvalues and eigenvectors of their respective normalized Laplacians, ensuring the permutation invariance and the unique description of each system. This property, in turn, is exploited to define metrics that quantify differences in chemo-spatial configurations near the binding site of an adsorbate. In particular, different atomic configurations will result in a different set of eigenvalues, which can be used to train a data-driven model to correct errors in the binding energies near the binding site.

Having uniquely determined differences between compositions using the previously presented procedure, we can now introduce the correction model. Using an input vector with the eigenvalues of ${\bf L}$, we can train the correction using a GPR. The primary advantage of GPR lies in its ability to offer a reasonable prediction power even with limited input data as it does not rely on fitting parameters to a predetermined basis function~\cite{banerjee2013efficient, deringer2021gaussian}. Instead, GPR infers correlations between the measured data points. Additionally, GPR can effectively relate multiple input variables to a single output variable, making it an ideal choice for our application~\cite{schulz2018tutorial}. Specifically, we use the eigenvalues resulting from the spatial and chemical distribution of the sample, with its APDFT BE prediction as input variables. GAP-DFT predicts the corrected BE of the binding site atoms, enabling the model to capture subtle relationships between the chemo-spatial environment and the APDFT errors. {\color{black}The implementation of the GPR model was done in Matlab~\cite{matlab}. Square exponential kernel, characterized by its smoothness and flexibility, was used~\cite{wilson2013gaussian}, 

\begin{equation} \label{eq:kernel}
k(\mathbf{x}_i, \mathbf{x}_j) = \sigma_f^2 \exp\left(-\frac{1}{2\ell^2} \|\mathbf{x}_i - \mathbf{x}_j\|^2\right)
\end{equation}
where $\sigma_f^2$ is the variance which determines the average distance of from the mean. The initial value for $\sigma_f$ was taken as the standard deviation of the data. $\ell$ is the characteristic length scale, with a value equal to the mean of the standard deviations of the predictors used in the model.}

\begin{figure}[H]
\centering
\includegraphics[trim={0cm  0.2cm 0cm 0cm },clip,width=0.9\columnwidth]{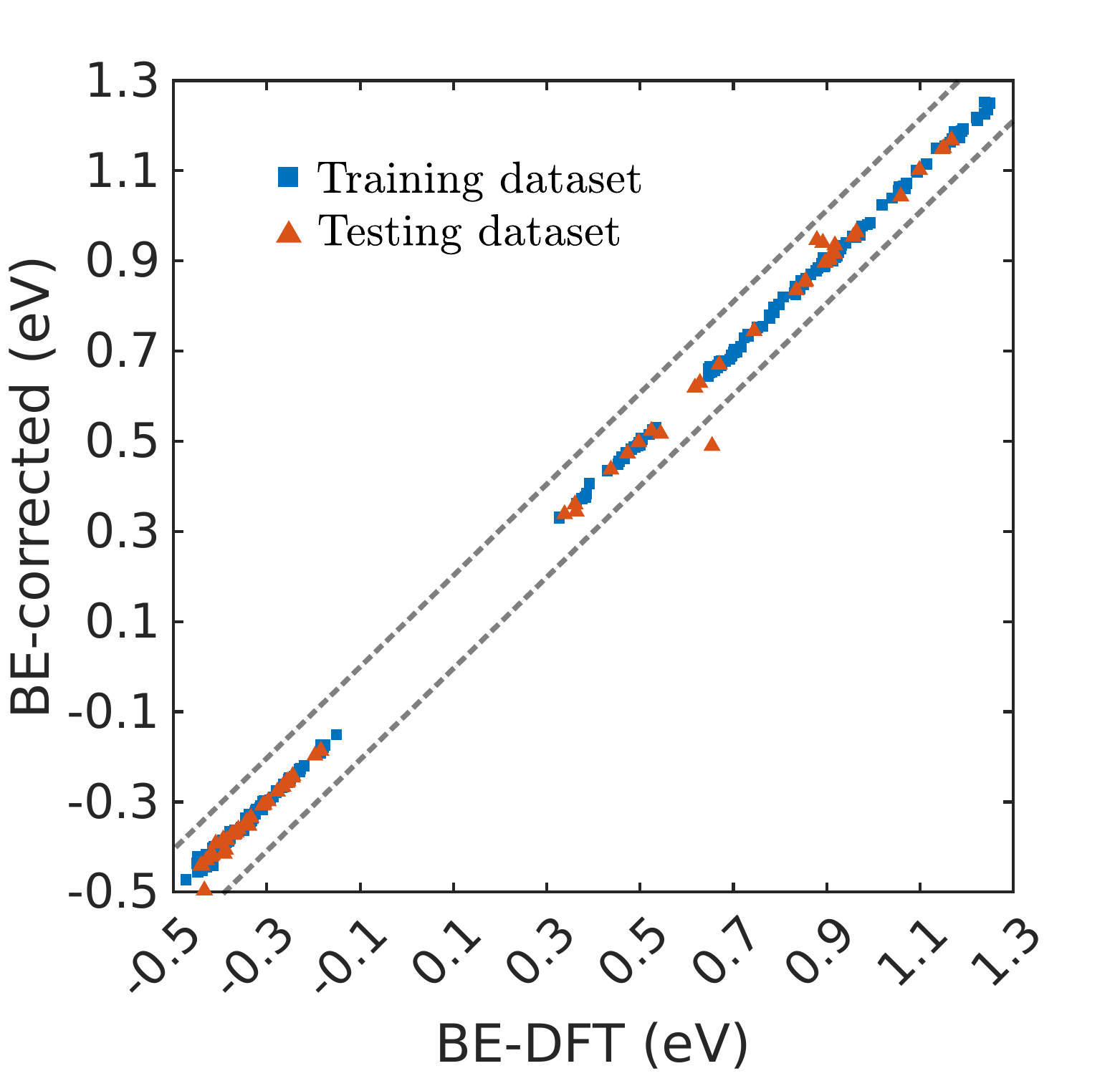}
\caption{The binding energy (BE) of the corrected APDFT using the Gaussian process regression (GPR) model compared to the BE for DFT.}
\label{Fig:gpr_val}
\end{figure}

To assess whether GAP-DFT would be a successful approach for correcting the BE values, we used the data for CO on top of Pd, Ag, Pt and Au, which had differences in the BE up to $0.6$~eV, as presented previously in Fig~\ref{Fig:HEA_conf_DFT_alch}. 
The input data, consisting of $324$ datapoints, has been divided randomly into $80\%$ training data and $20\%$ testing data. The results of the predicted BE using the GAP-DFT model and the DFT for the training and testing datasets are shown in Fig.~\ref{Fig:gpr_val}. The predicted BE values by GAP-DFT were close to the DFT reference values.

\begin{figure}[H]
\centering
\includegraphics[trim={0cm  0.2cm 0cm 0cm },clip,width=0.99\columnwidth]{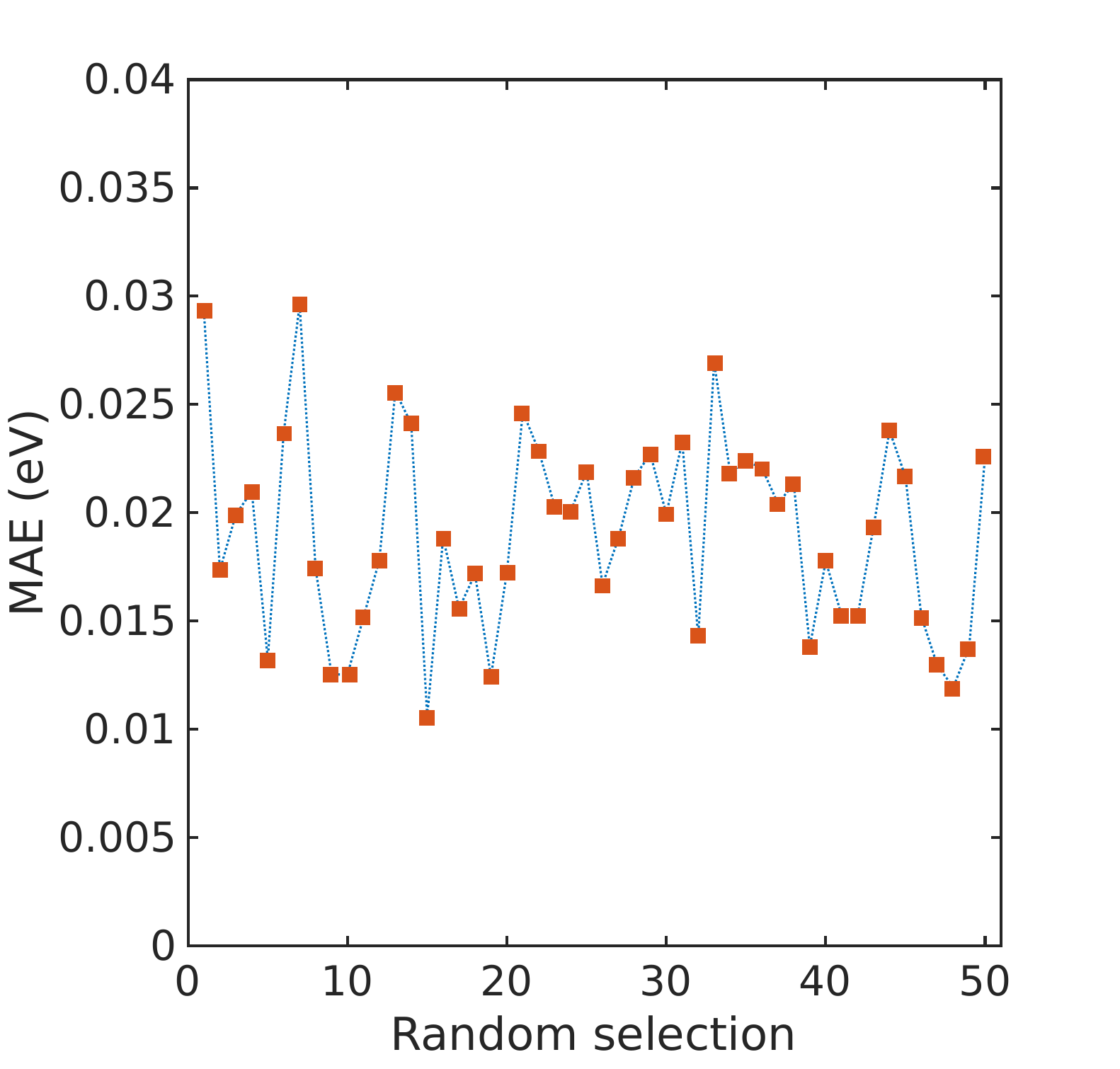}
\caption{Mean average error using the GAP-DFT. Each random selection means the training set was randomly selected for each instance. For this case, 80~\% of data points were used for training, the remaining for testing purposes.
}
\label{Fig:gpr_err}
\end{figure}

To make sure that the prediction accuracy is independent of the split of the training and testing datasets, we randomly generated $50$ different training-testing datasets. The MAE was reported for the $50$ random selections, as shown in Fig.~\ref{Fig:gpr_err}. The MAE obtained with GAP-DFT  for the testing set was reported between $0.01$~eV and $0.03$~eV, which is reasonable and reduces the high error of around $0.6$~eV of the APDFT predictions.

\subsection*{GAP-DFT binding energies distribution}
 We investigated the BE distribution after introducing the corrected GAP-DFT approach for HEAs. Unlike pure elements catalysts with a single BE, HEAs have a BE distribution, which depends on the binding site element(s) and the surrounding chemical environment~\cite{batchelor2019high}. 

\begin{figure*}
\centering
\subfloat[\centering DFT 20 simulations]{\includegraphics[trim={0cm  0cm 0cm 3cm }, width=0.33\textwidth]{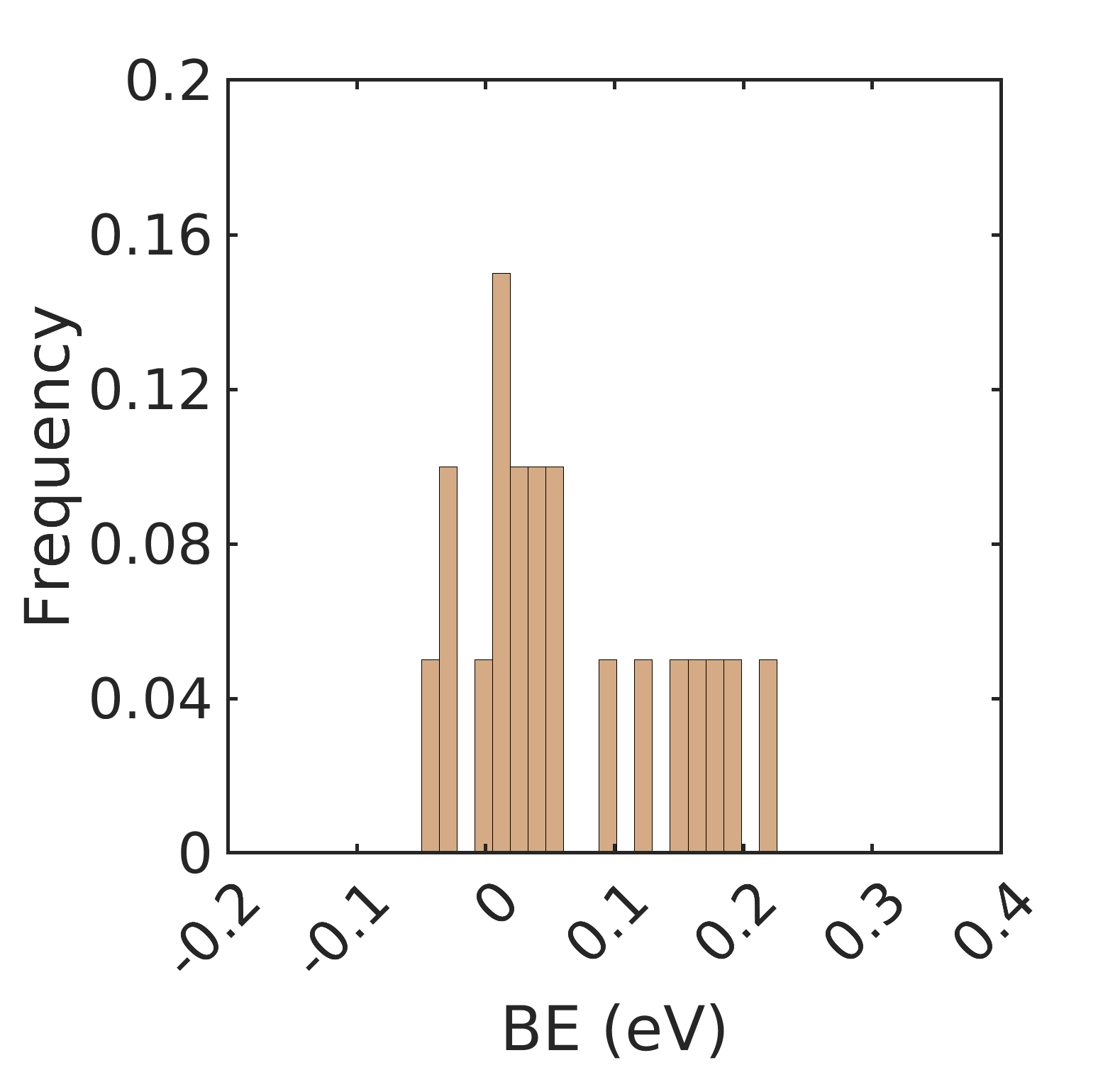}}
\subfloat[\centering GAP-DFT 20 simulations, 1-permutation]{\includegraphics[trim={0cm  0cm 0cm 3cm }, width=0.33\textwidth]{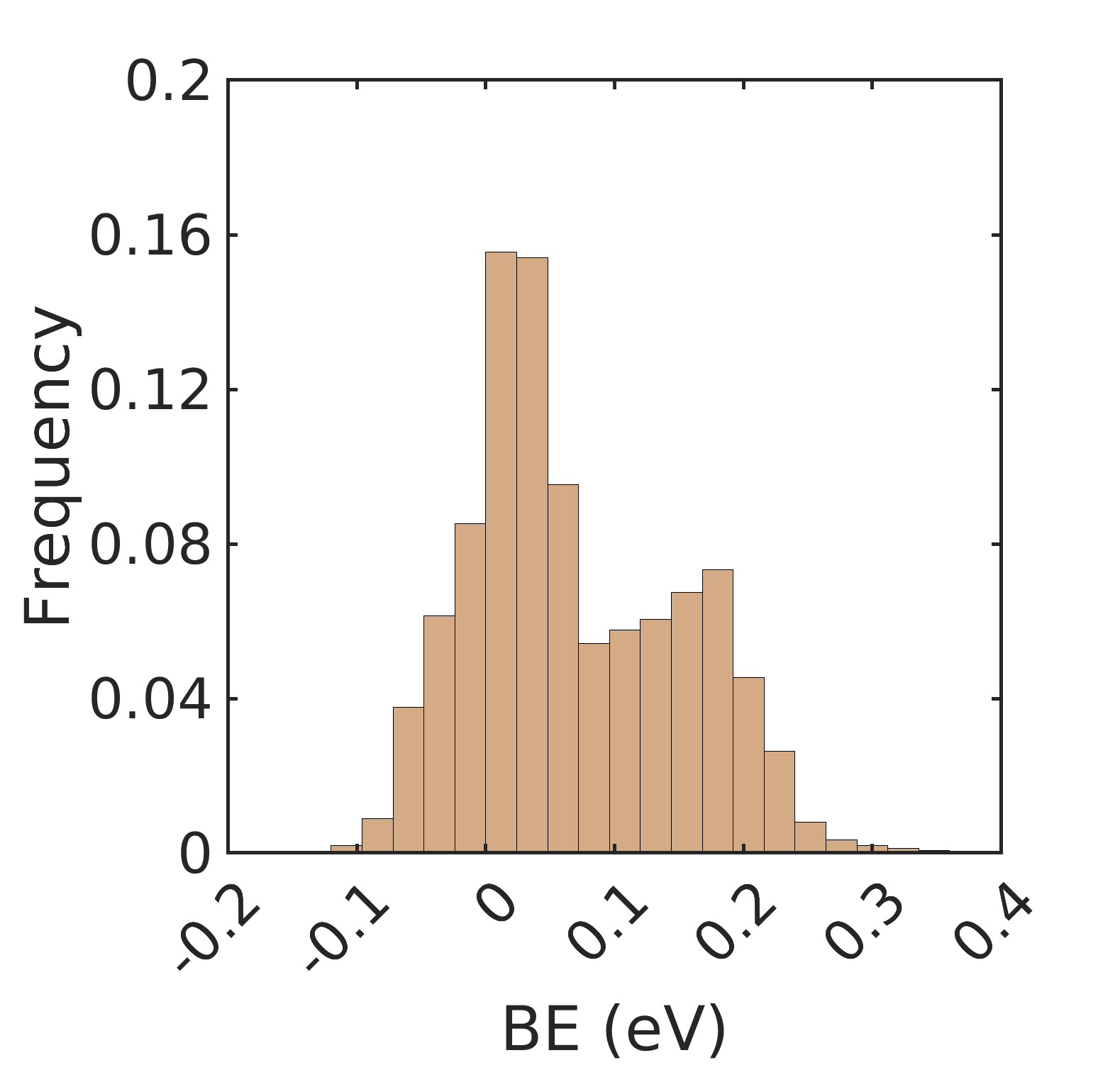}}
\subfloat[\centering GAP-DFT 20 simulations, 2-permutations]{\includegraphics[trim={0cm  0cm 0cm 3cm }, width=0.33\textwidth]{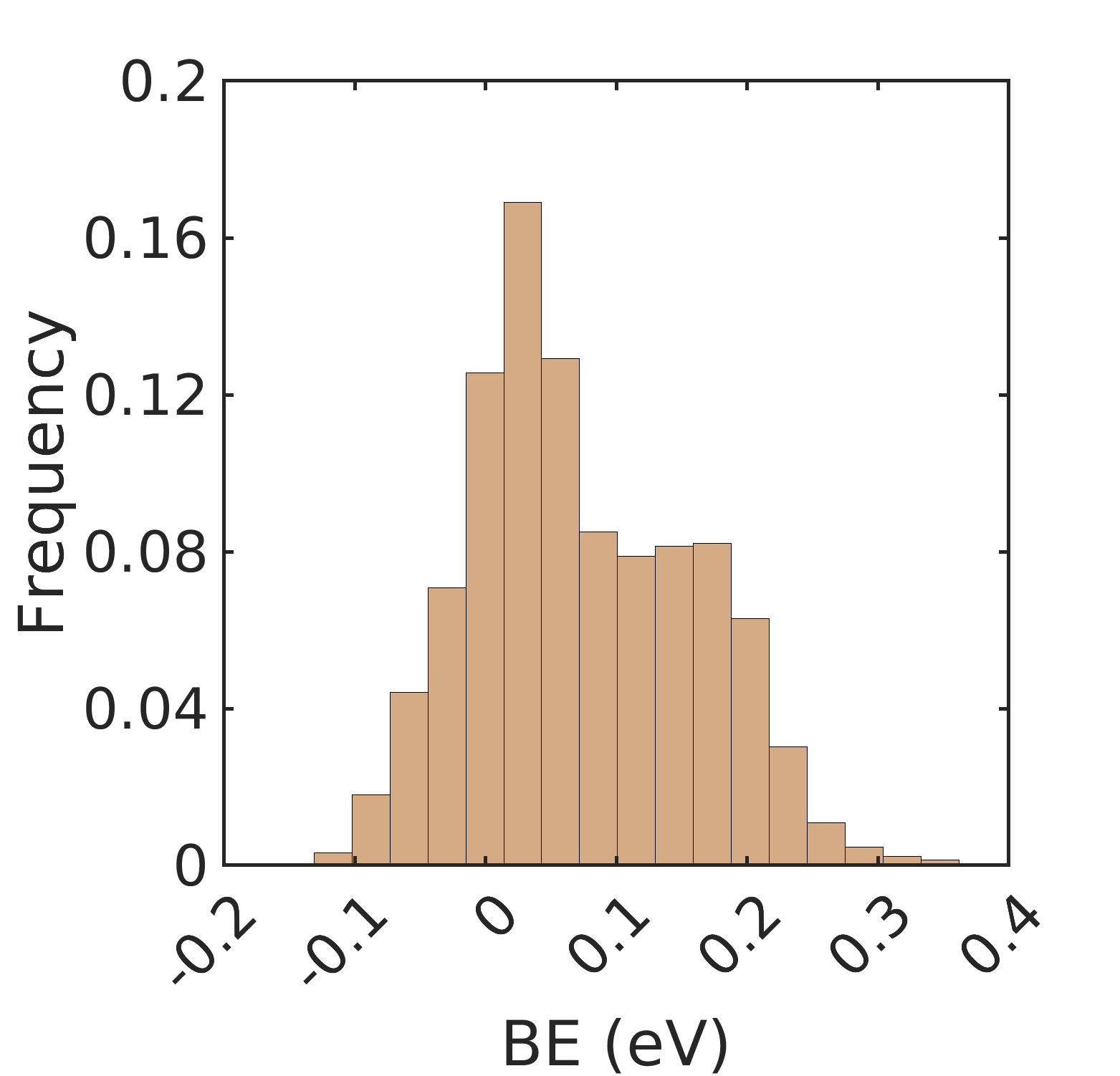}}
\bigbreak
\subfloat[\centering DFT 100 simulations]{\includegraphics[trim={0cm  0cm 0cm 3cm }, width=0.33\textwidth]{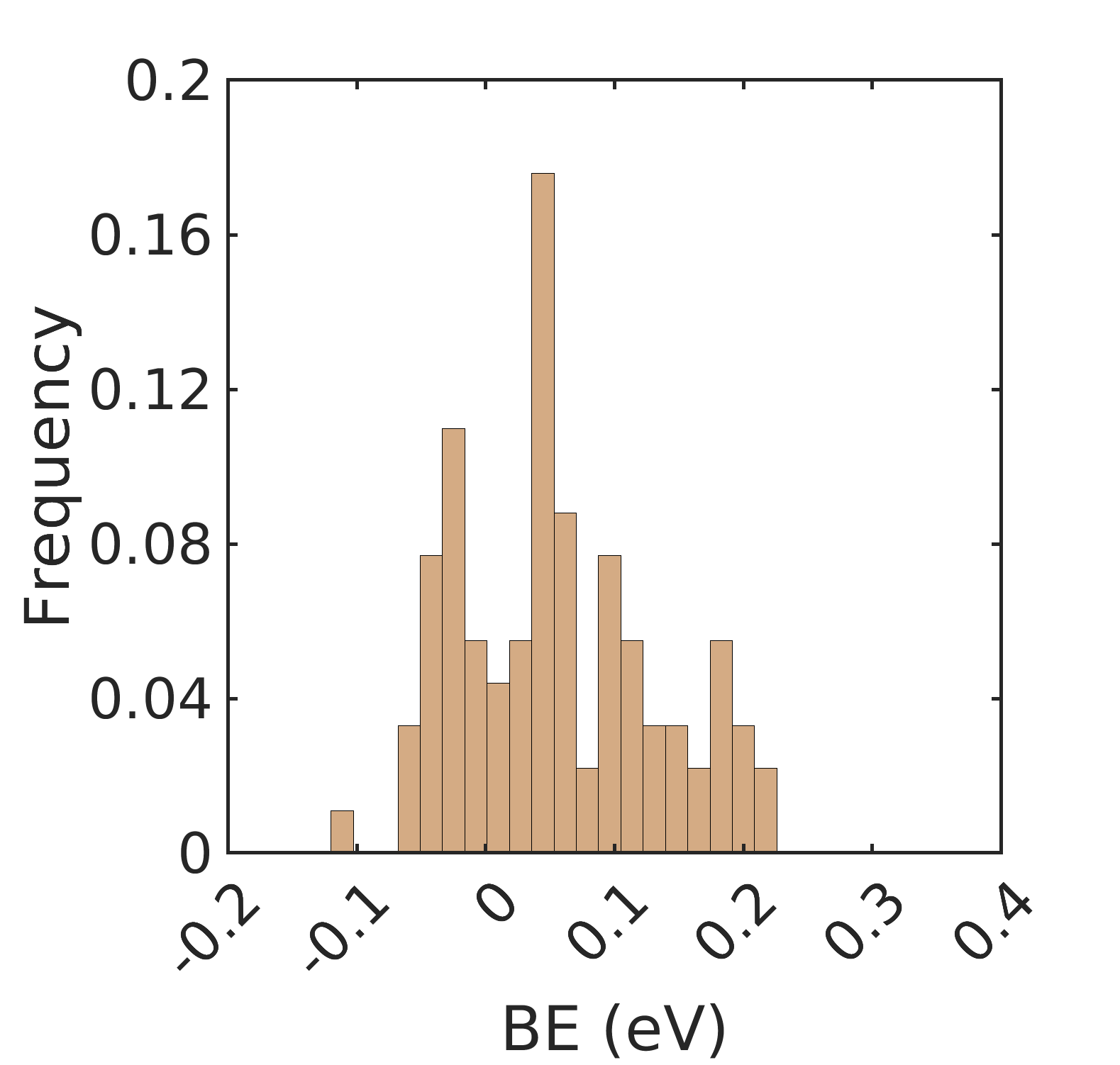}}
\subfloat[\centering GAP-DFT 100 simulations, 1-permutation]{\includegraphics[trim={0cm  0cm 0cm 3cm }, width=0.33\textwidth]{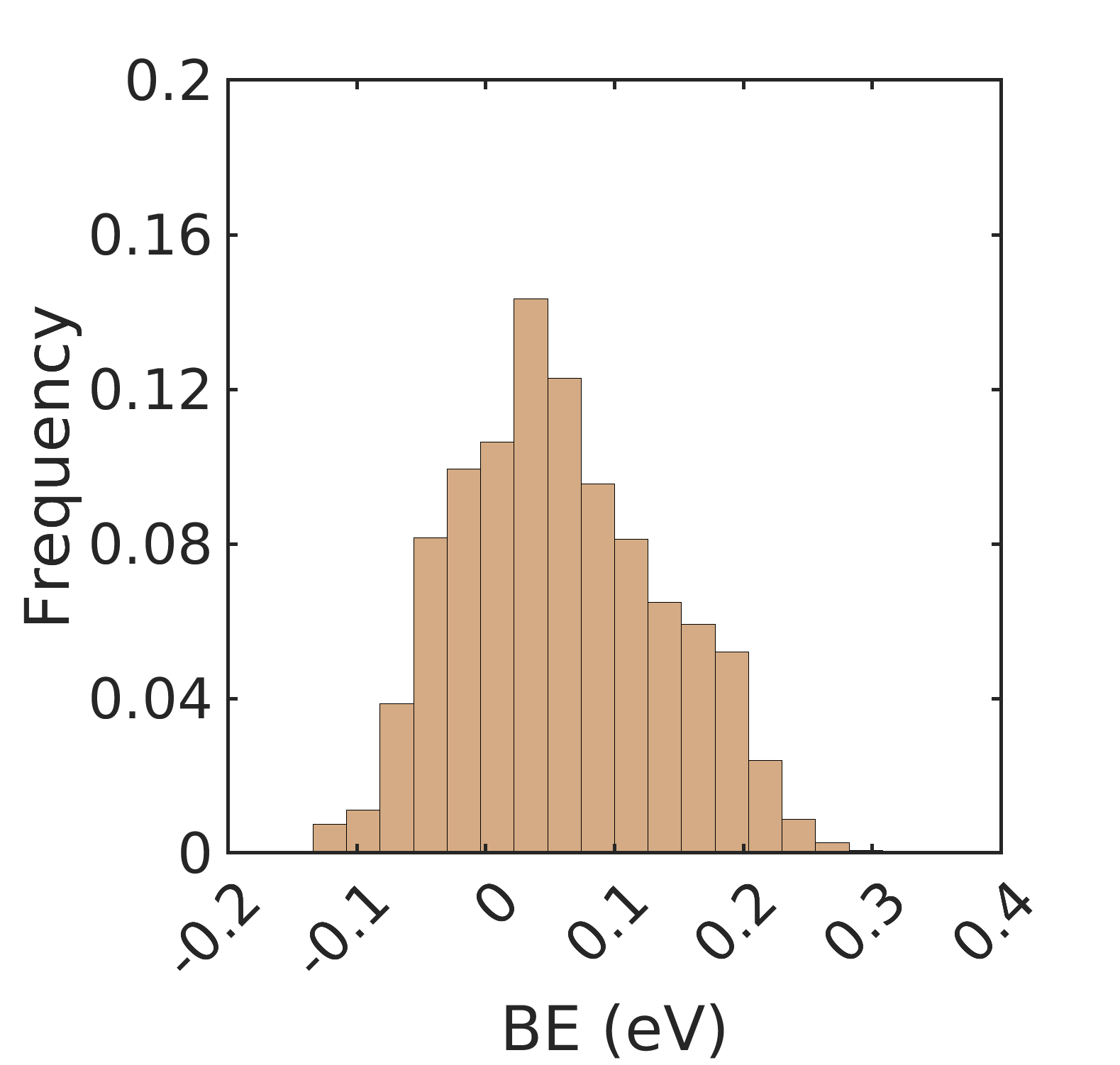}}
\subfloat[\centering GAP-DFT 100 simulations, 2-permutations]{\includegraphics[trim={0cm  0cm 0cm 3cm }, width=0.33\textwidth]{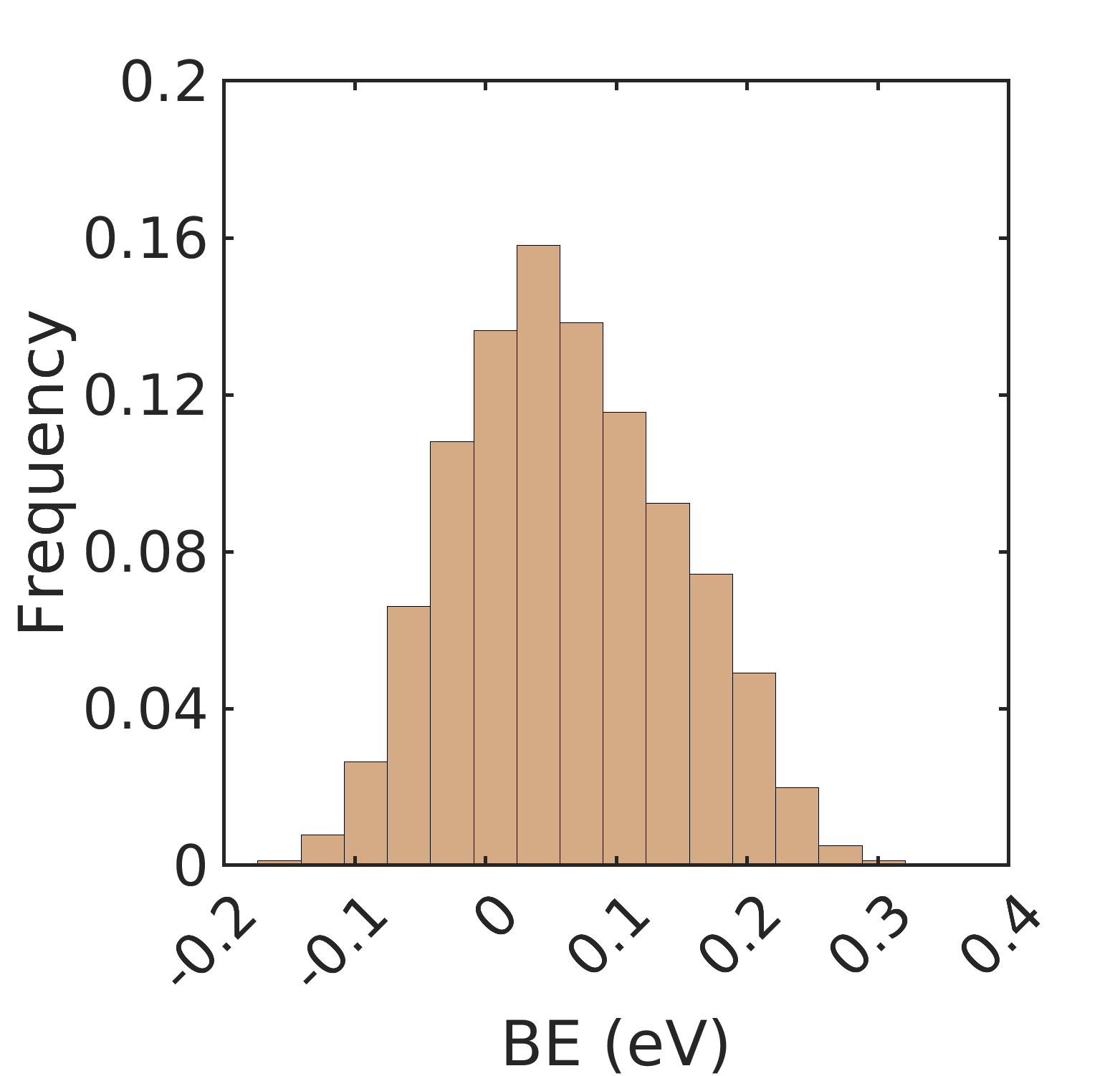}}
\bigbreak
\caption{The binding energy (BE) distribution for CO on-top binding site of Cu using (a) 20 DFT calculations, (b) GAP-DFT with 20 DFT and one permutation at a time, (c) GAP-DFT with 20 DFT and two permutations at a time, (d) 100 DFT calculations, (e) GAP-DFT with 100 DFT and one permutation at a time, (f) GAP-DFT with 100 DFT and two permutations at a time.
}
\label{Fig:histo_hea}
\end{figure*}

We aim to obtain the BE distributions by utilizing the minimum number of DFT calculations and their corresponding $-$and computationally inexpensive$-$ GAP-DFT calculations. 
The BE distributions can be useful in screening and optimizing the HEA composition for catalytic applications.
The histogram of the BE of $20$ DFT calculations of CO on-top of Cu is shown in Fig.~\ref{Fig:histo_hea}(a). This number of DFT calculations is extremely small and cannot represent the BE distribution accurately. It is worth mentioning that the range of BE is between $-0.04$~eV to $0.22$~eV. Interestingly, using 20 DFT simulations as input for GAP-DFT with one permutation at a time allowed us to predict a wider range of BE (i.e., $-0.12$~eV to $0.36$~eV), drastically populating the BE distribution, as depicted in Fig.~\ref{Fig:histo_hea}(b). Including two simultaneous permutations (see Fig.~\ref{Fig:histo_hea}(c)) slightly changed the BE distribution, indicating more coverage of the compositional space as the BE is affected by the element in the binding site and the surrounding atoms.

To further assess if the corrected APDFT method predicts the right trends, we performed calculations using 100 DFT simulations with CO on-top of Cu, shown in  Fig.~\ref{Fig:histo_hea}(d). Remarkably, the lower bound of the BE distribution is the same as the one predicted with the APDFT method. We point out that even with this number of DFT simulations, the BE distribution is poorly represented and does not capture all the subtleties of the distribution. Performing one and two swaps improves the BE distribution, similar to the 20 DFT cases, as illustrated in Figs.~\ref{Fig:histo_hea}(e) and (f).

To quantify the convergence of the distributions using different numbers of simulations and APDFT extensions, we introduce the Kullback-Leibler (KL) divergence to measure the similarity between each distribution. 
The KL divergence, often called relative entropy, is an essential metric in information theory and statistics to quantify the divergence between two probability distributions. 
In this study, the KL divergence and the area intersection methods are utilized to compare the histograms of the DFT calculations and their corresponding APDFT estimations for non-binding site atoms, providing a rigorous measure of how one probability distribution \(P\) diverges from a second reference distribution \(Q\). Mathematically, the KL divergence is defined as

{\color{black}
\begin{equation} \label{eq:KL1}
{D}_\mathrm{KL}(P \parallel Q) = \sum_i P(i) \log \frac{P(i)}{Q(i)},
\end{equation}
}

where \(P(i)\) and \(Q(i)\) represent the probability values of the \(i\)-th bin in the respective histograms. A KL divergence value of zero indicates perfect alignment between the two distributions, while higher values denote increasing levels of dissimilarity.

\begin{figure}[h]
\centering
\subfloat[Area intersection fraction]{\includegraphics[trim={0cm  0cm 0cm 0cm }, width=0.9\columnwidth]{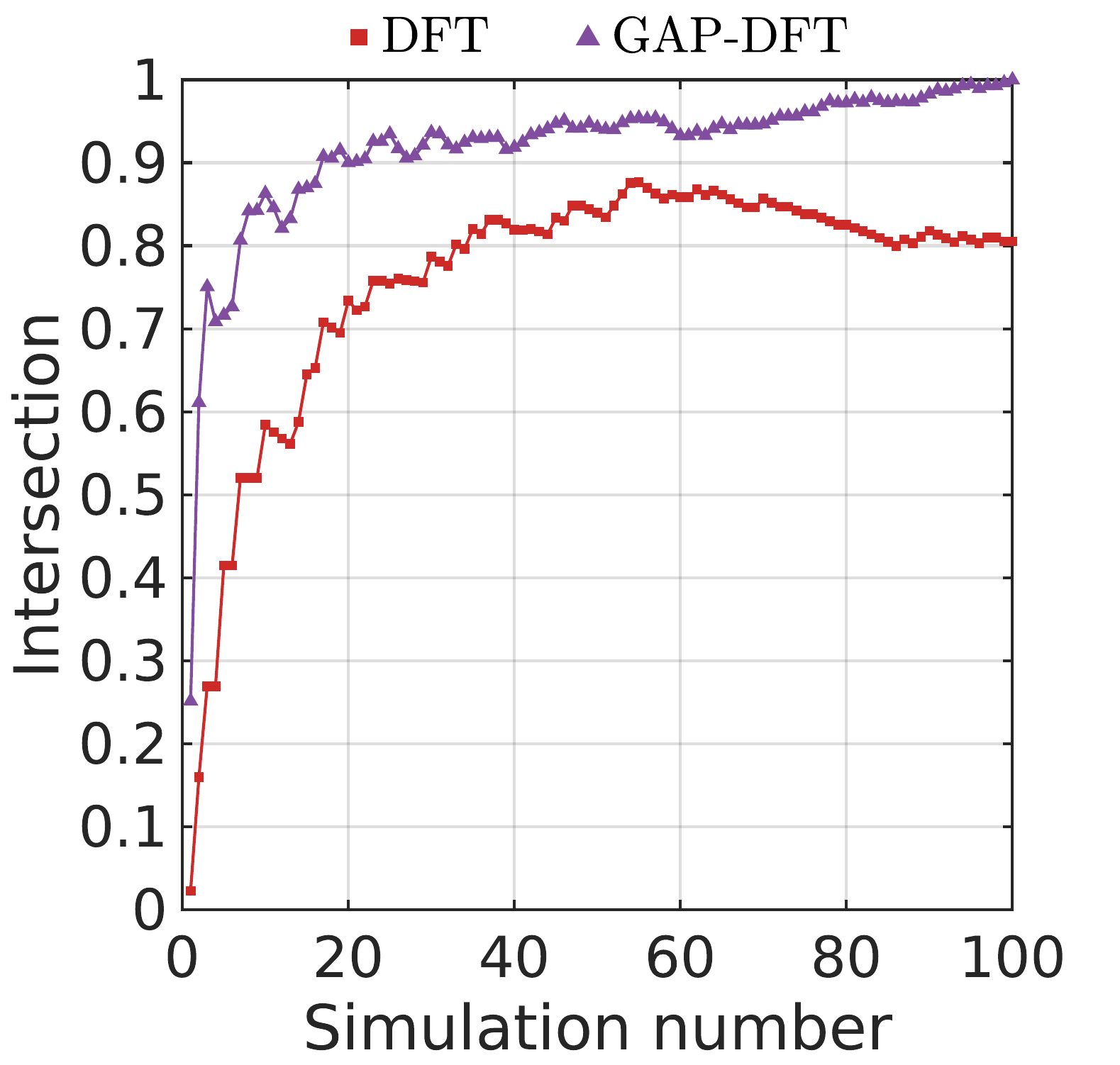}}\\
\subfloat[KL divergence]{\includegraphics[trim={0cm  0cm 0cm 0cm },clip,width=0.9\columnwidth]{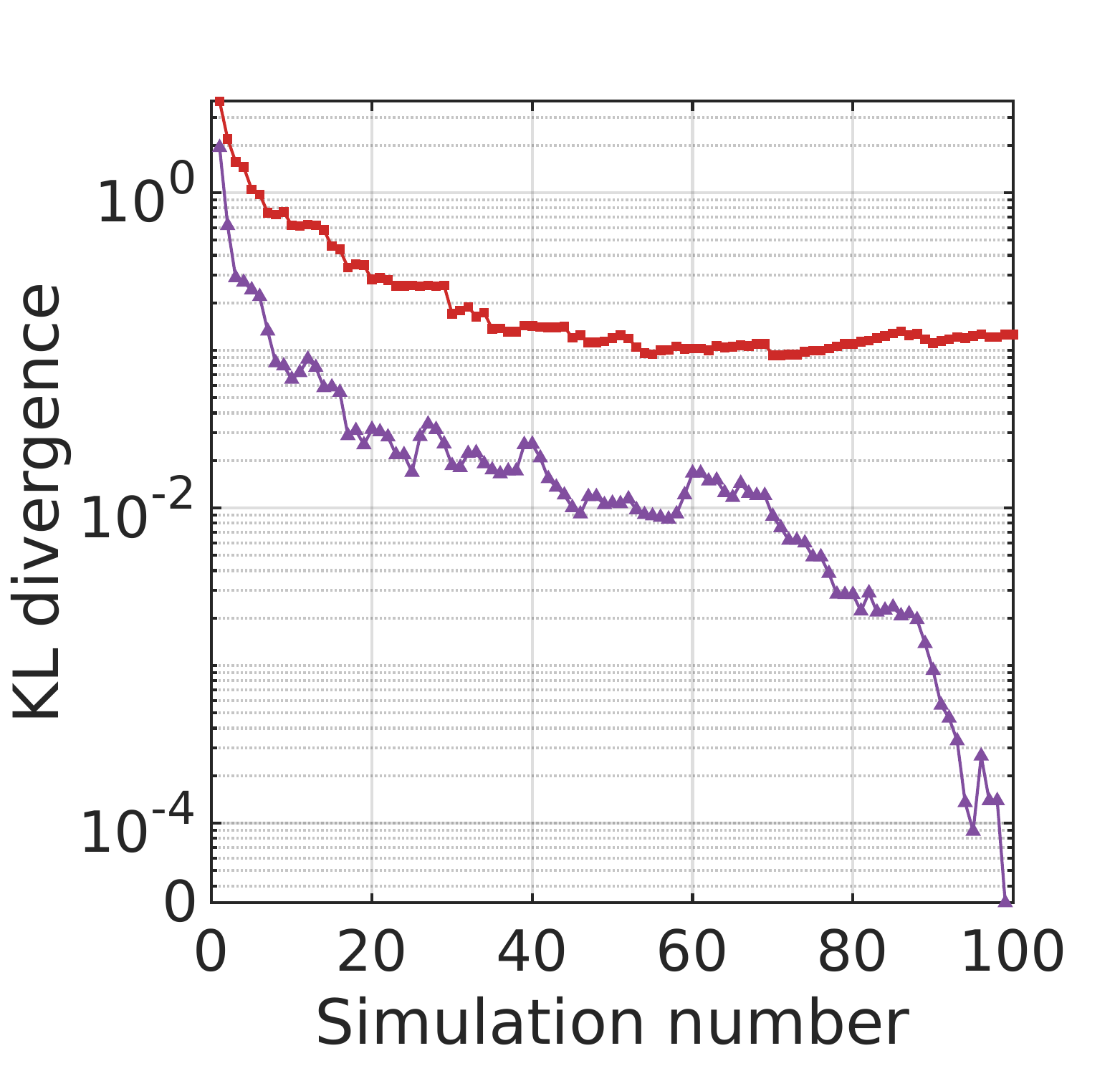}
}\hfill
\caption{(a) Comparison between DFT and GAP-DFT using the intersection method. (b) Compared with the reference distribution, the Kullback-Leibler (KL) divergence between cumulative DFT distributions and GAP-DFT distributions. 
}
\label{Fig:KL}
\end{figure}

Figs.~\ref{Fig:KL}(a) and (b) show the area intersection and the KL divergence for DFT and GAP-DFT methods, respectively. The histogram built with GAP-DFT using 100 DFT simulations and two permutations is used as a reference to compute the intersection and KL divergence. We notice that the DFT simulations cover ~70\% ($D_{KL}=0.3$) of the total area with ~20 simulations, whereas the GAP-DFT covers ~90\% (${D}_\mathrm{KL}=0.03$) for the same number of DFT simulations. Furthermore, DFT simulations alone cannot cover the same area as APDFT, reaching about 80\% when 100 DFT simulations were used. This is also denoted by the KL divergence, which cannot be reduced below $\sim0.1$. The increase in the overlap area between 50-80 DFT simulations can be attributed to variations in the histogram due to the bin size that changes the relative area of the distribution. However, ${D}_\mathrm{KL}$ remains around $0.1$ for the same range, indicating no quantitative changes in the distributions. ADPFT increases the intersection in an almost linear trend, with 95\% overlap when 50-60 DFT simulations were used (${D}_\mathrm{KL}=0.01$). 

\begin{figure}[h]
\centering
\includegraphics[width=\columnwidth]{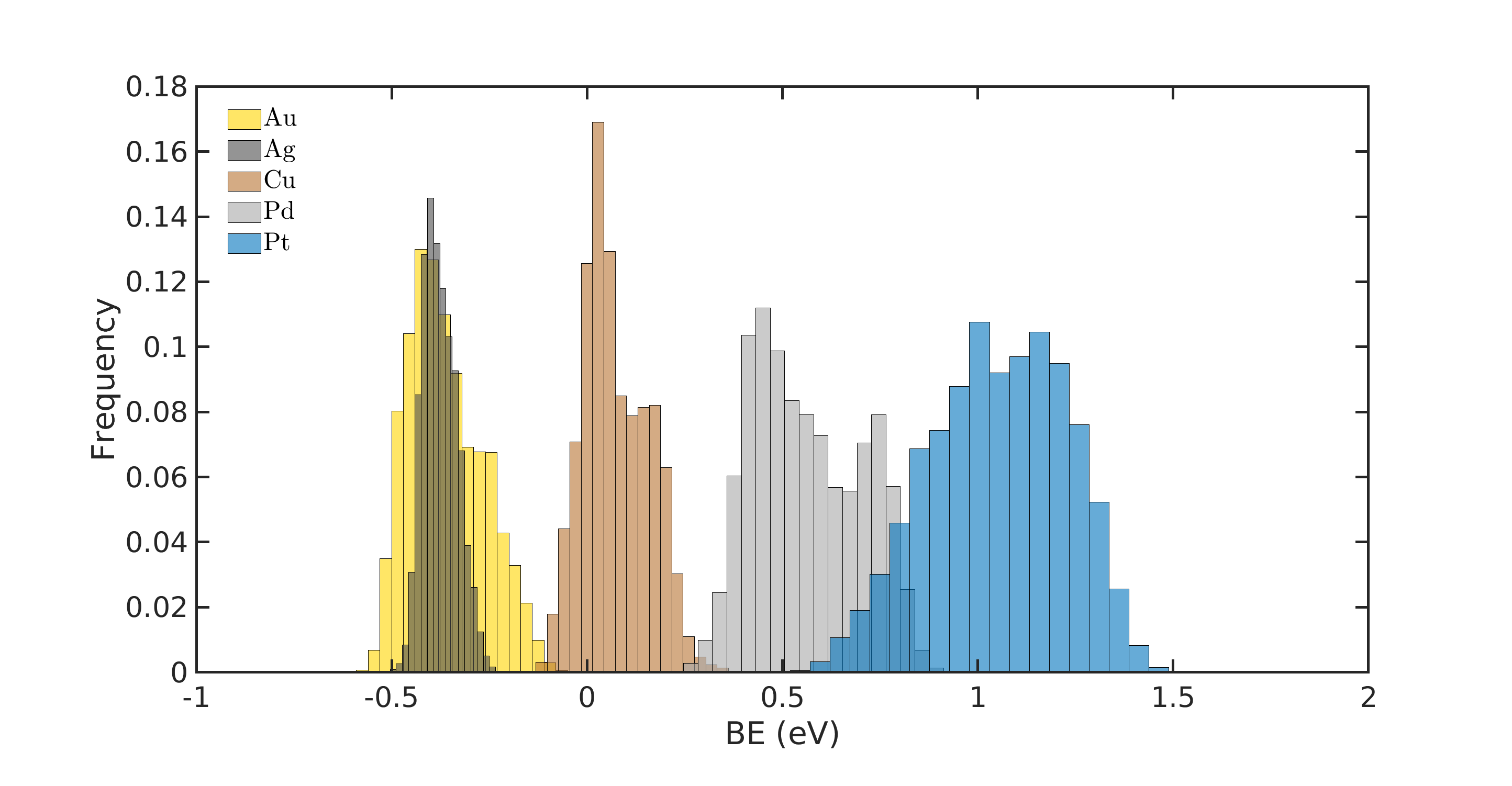}
\caption{The Binding energy (BE) distribution of CO on-top site over the elements of CuPdAgPtAu HEA based on the introduced framework. The distributions have been computed with GAP-DFT using 100 DFT simulations (20 for each element) with one and two permutations.}
\label{Fig:BE_dist}
\end{figure}

Based on the finding above, the BE distributions can be built using GAP-DFT with just $20$ DFT calculations and two permutations. Hence, $20$ DFT calculations of CO on-top of each one of the five elements in the HEA were utilized to plot the BE distribution for the CuPdAgPtAu HEA, which is shown in Fig.~\ref{Fig:BE_dist}. Remarkably, the BE distributions involve $\sim664,200$ simulations, showcasing the tremendous computational power of GAP-DFT. The BE shows that Ag and Au are weak adsorbents for CO, while Pd and Pt are strong adsorbents {\color{black}in this HEA, similar to another work~\cite{pedersen2020high}}. Interestingly, Cu, the most active catalyst towards higher carbon products, has, on average, slightly strong adsorption to CO. This agrees with the previous results in the literature. For instance, in their study on the same equimolar HEA, Pedersen \emph{et al.} predicted a distribution of BE in agreement with Fig.~\ref{Fig:BE_dist}, with a range of BE between $-0.2$~eV to $1.6$~eV which is good agreement with our ranges between $-0.52$~eV to $1.5$~eV~\cite{pedersen2020high}. The wider range found in our case can be attributed to the vast sampling of the BE distribution performed by APDFT. 

Furthermore, the predicted BE distribution by APDFT for CO on-top of Pd and Pt is broad compared to Cu, Ag and Au, whereas there is a significant overlap for the BE distributions for Ag and Au. These observations agree well with previous results~\cite{pedersen2020high}. It is worth mentioning that the results in~\cite{pedersen2020high} are based on an ML model built using $\sim2000$ DFT data points. In contrast, our BE distributions are based on just $100$ DFT calculations and their corresponding APDFT estimations, which are calculated analytically without additional computational cost. This shows the significant computational efficiency of our approach, offering at least one order of magnitude acceleration compared to other approaches.   

\section{Conclusion}\label{sec3}
We applied the alchemy perturbation density functional method to high-entropy alloys for catalytic applications. We have shown that the binding energies for sites away from the binding sites can be predicted accurately with computational alchemy without incurring significant computational costs. However, the BE errors are exceedingly large for binding sites atom(s). To remedy this, we presented GAP-DFT, a graph-based correction scheme using the Gaussian regression method to predict binding energies for binding sites. We showed that the GAP-DFT can predict mean average errors of about 30 meV, and used it to build the BE distributions for an equiatomic CuPdAgPtAu HEA, focusing on the CO molecule. We showed that using GAP-DFT with only 20 DFT simulations, we constructed distributions that cover 90\% of the BE space. Our results are in both qualitative and quantitative agreement with previously reported works, but at a computational cost that is an order of magnitude lower than the \emph{state-of-the-art} data-driven methods. {\color{black}By significantly reducing computational costs, the GAP-DFT can accelerate the design and optimization of new HEAs for catalytic applications. This approach holds great promise for overcoming current challenges in energy technology by enabling the rapid exploration of vast compositional spaces, thereby facilitating the discovery of more efficient and sustainable catalytic materials.}


\section*{Author contributions}
Mohamed Hendy: Methodology, Software, Data Curation, Formal analysis, Investigation, Visualization and Writing - Original Draft. Okan K. Orhan: Conceptualization, Methodology and Writing - Review \& Editing. Homin Shin: Conceptualization, Methodology and Writing - Review \& Editing. Ali Malek: Conceptualization, Methodology and Writing - Review \& Editing. Mauricio Ponga: Conceptualization, Methodology, Resources, Writing - Review \& Editing, Supervision, Project administration and Funding acquisition.

\section*{Conflicts of interest}
The authors declare no conflicts of interest.

\section*{Data availability}
The codes developed during this study and the data used are available at our Github directory~\cite{GAPDFTProject}.

\section*{Acknowledgements}

We acknowledge the support of the National Research Council Canada via Contract No. 987243, New Frontiers in Research Fund (NFRFE-2019-01095) and the Natural Sciences and Engineering Research Council of Canada (NSERC) through the Discovery Grant.  M.H. gratefully acknowledges the financial support from the Department of Mechanical Engineering at UBC through the Four Years Fellowship. 
This research was supported through high-performance computational resources and services provided by Advanced Research Computing at the University of British Columbia and the Digital Research Alliance of Canada. 

\bibliography{rsc}

\end{document}